%% file: ms.tex
\documentclass[preprint]{aastex}

\usepackage{natbib}

\newcommand{\figscale}{1.0}
\newcommand{\thisdir}{.}
\newcounter{tmp_eqn_count}
\newcounter{ref_count}

\begin{document}

	\include{\thisdir/H2_macros}

	\include{\thisdir/paperabstract}

	\include{\thisdir/intro}
	\include{\thisdir/chemistry}
	\include{\thisdir/numerics}

	\include{\thisdir/collapse}

	\include{\thisdir/cosmo}
	\include{\thisdir/summary}
	\include{\thisdir/bibliog}

\end{document}

%% file: H2_macros.tex
	\newlength{\fntxvi} 
	\newlength{\fntxvii}
	\newcommand{\chemical}[1]
	{{\fontencoding{OMS}\fontfamily{cmsy}\selectfont
	  \fntxvi\the\fontdimen16\font
	  \fntxvii\the\fontdimen17\font
	  \fontdimen16\font=3pt\fontdimen17\font=3pt
	  \ensuremath{\mathrm{#1}}
	  \fontencoding{OMS}\fontfamily{cmsy}\selectfont
	  \fontdimen16\font=\fntxvi \fontdimen17\font=\fntxvii}}

	\newcommand{\Hn}{\chemical{H}}
	\newcommand{\Hm}{\chemical{H^-}}
	\newcommand{\Hp}{\chemical{H^+}}
	\newcommand{\Ht}{\chemical{H_2}}
	\newcommand{\Htp}{\chemical{H_2^+}}
	\newcommand{\HTp}{\chemical{H_3^+}}
	\newcommand{\el}{\chemical{e^-}}
	\newcommand{\HD}{\chemical{HD}}
	\newcommand{\LiH}{\chemical{LiH}}
	\newcommand{\He}{\chemical{He}}
	\newcommand{\Hep}{\chemical{He^+}}
	\newcommand{\Hepp}{\chemical{He^{++}}}
	
	\newcommand{\fHn}{\ensuremath{f_{\Hn}}}
	\newcommand{\fHm}{\ensuremath{f_{\Hm}}}
	\newcommand{\fHp}{\ensuremath{f_{\Hp}}}
	\newcommand{\fHt}{\ensuremath{f_{\Ht}}}
	\newcommand{\fHtp}{\ensuremath{f_{\Htp}}}
	\newcommand{\fHTp}{\ensuremath{f_{\HTp}}}
	\newcommand{\fel}{\ensuremath{f_{\el}}}
	\newcommand{\fHD}{\ensuremath{f_{\HD}}}
	\newcommand{\fLiH}{\ensuremath{f_{\LiH}}}
	\newcommand{\fHe}{\ensuremath{f_{\He}}}
	\newcommand{\fHep}{\ensuremath{f_{\Hep}}}
	\newcommand{\fHepp}{\ensuremath{f_{\Hepp}}}
	
	\newcommand{\Li}{\chemical{Li}}
	\newcommand{\D}{\chemical{D}}
	\newcommand{\Het}{\chemical{^3He}}
	\newcommand{\Hef}{\chemical{^4He}}

  \newcommand{\Msun}{\ensuremath{\rm{M}_\odot}}
	\newcommand{\Lya}{\ensuremath{\textrm{Ly}\alpha}}

\newcommand{\rHpelHng}{\ensuremath{k_{\ref{HpelHng}}}}
\newcommand{\rHnelHmg}{\ensuremath{k_{\ref{HnelHmg}}}}
\newcommand{\rHnelHpTel}{\ensuremath{k_{\ref{HnelHpTel}}}}
\newcommand{\rHmelHnTel}{\ensuremath{k_{\ref{HmelHnTel}}}}
\newcommand{\rHmHpTHn}{\ensuremath{k_{\ref{HmHpTHn}}}}
\newcommand{\rHmHnTHnel}{\ensuremath{k_{\ref{HmHnTHnel}}}}
\newcommand{\rHnHnHt}{\ensuremath{k_{\ref{HnHnHt}}}}
\newcommand{\rHmHnHtel}{\ensuremath{k_{\ref{HmHnHtel}}}}
\newcommand{\rHtpHnHtHp}{\ensuremath{k_{\ref{HtpHnHtHp}}}}
\newcommand{\rHtpHmHtHn}{\ensuremath{k_{\ref{HtpHmHtHn}}}}
\newcommand{\rHTpelHtHn}{\ensuremath{k_{\ref{HTpelHtHn}}}}
\newcommand{\rHtHmHtHnel}{\ensuremath{k_{\ref{HtHmHtHnel}}}}
\newcommand{\rHtelTHnel}{\ensuremath{k_{\ref{HtelTHnel}}}}
\newcommand{\rHtelHnHm}{\ensuremath{k_{\ref{HtelHnHm}}}}
\newcommand{\rHtHpHtpHn}{\ensuremath{k_{\ref{HtHpHtpHn}}}}
\newcommand{\rHtHnThHn}{\ensuremath{k_{\ref{HtHnThHn}}}}
\newcommand{\rHtHtHtHn}{\ensuremath{k_{\ref{HtHtHtHn}}}}
\newcommand{\rHtHtpHTpHn}{\ensuremath{k_{\ref{HtHtpHTpHn}}}}
\newcommand{\rHpHnHtpg}{\ensuremath{k_{\ref{HpHnHtpg}}}}
\newcommand{\rHmHpHtpel}{\ensuremath{k_{\ref{HmHpHtpel}}}}
\newcommand{\rHtpelTHn}{\ensuremath{k_{\ref{HtpelTHn}}}}
\newcommand{\rHngHpel}{\ensuremath{k_{\ref{HngHpel}}}}
\newcommand{\rHmgHnel}{\ensuremath{k_{\ref{HmgHnel}}}}
\newcommand{\rHtpgHnHp}{\ensuremath{k_{\ref{HtpgHnHp}}}}
\newcommand{\rHtpgTHpel}{\ensuremath{k_{\ref{HtpgTHpel}}}}
\newcommand{\rHtgHtpel}{\ensuremath{k_{\ref{HtgHtpel}}}}
\newcommand{\rHtgHpHnel}{\ensuremath{k_{\ref{HtgHpHnel}}}}
\newcommand{\rHtgHnHn}{\ensuremath{k_{\ref{HtgHnHn}}}}
\newcommand{\rHtgTHn}{\ensuremath{k_{\ref{HtgTHn}}}}

\newcommand{\nbody}{$N$-body }

\newcommand{\Kelvin}{\ensuremath{\,\mathrm{K}}}

%% file: paperabstract.tex
\title{Simulation of primordial object formation}
\author{Todd M. Fuller and H. M. P. Couchman\altaffilmark{1}} 
\email{tfuller@astro.uwo.ca}
\email{couchman@mcmaster.ca}

\affil{Department of Physics and Astronomy, University of Western Ontario\\
London, ON N6A~3K7, Canada}

\input{\thisdir/abstract}

\altaffiltext{1}{Department of Physics and Astronomy, McMaster University, Hamilton, ON L8S~4M1, Canada}


%% file: abstract.tex
\begin{abstract}

We have included the chemical rate network responsible for the formation of
molecular Hydrogen in the $N$-body hydrodynamic code, Hydra, in order to study
the formation of the first cosmological at redshifts between 10 and 50.   We
have tested our implementation of the chemical and cooling processes by
comparing N-body top hat simulations with theoretical predictions from a
semi-analytic model and found them to be in good agreement. We find that
post-virialization properties are insensitive to the initial abundance of
{\Ht}. Our main objective was to determine the minimum mass ($M_{SG}(z)$) of
perturbations that could become self gravitating (a prerequisite for star
formation), and the redshift at which this occurred. We have developed a robust
indicator for detecting the presence of a self-gravitating cloud in our
simulations and find that we can do so with a baryonic particle mass-resolution
of 40\Msun. We have performed cosmological simulations of primordial objects
and find that the object's mass and redshift at which they become self
gravitating agree well with the $M_{SG}(z)$ results from the top hat
simulations.  Once a critical {\Ht} fractional abundance of $\sim 5 \times
10^{-4}$ has formed in an object, the cooling time drops below the dynamical
time at the centre of the cloud and the gas free falls in the dark matter
potential wells, becoming self gravitating a dynamical time later.

\end{abstract}

%% file: intro.tex
\section{Introduction}

In cold dark matter (CDM) cosmologies, the first objects are expected to form
between redshifts of 10 and 50 \citep{Peebles93}.  Radiative cooling via the
rotational and vibrational transitions of molecular hydrogen provides the
mechanism by   which the baryons may first condense in small dark matter
potential wells of mass  $\sim 10^6\,\Msun$ and virial temperature $\sim
1000\Kelvin$; too cool for hydrogen line cooling to be effective.  Of the
several different molecules present in the early universe --- for example
{\Ht}, {\HD}, and {\LiH} --- molecular  hydrogen is the most abundant and
dominates cooling between $100\Kelvin$ and $1000\Kelvin$. Understanding
the cooling and formation of the first objects and their influence on
subsequent cosmic evolution is an important goal of post-recombination
cosmography. Of particular interest is the variety of feedback mechanisms which
may operate following the epoch of first star formation. 

Even a small number of objects at this early time could have a significant
effect on the the entire universe.  A fraction of just $\sim 10^{-5}$ of the
baryonic mass condensing into massive stars, would produce sufficient energetic
photons to re-ionize the universe \citep{Carr84,Couchman86,Loeb97}. 
Observations of the CMB provide a limit on the earliest time that reionization
could occur.  Anisotropies in the CMB on angular scales below the horizon size
of $\sim 10^\circ$ are diminished by free electron production during
reionization, and if reionization occurred much before $z \simeq 50$, the level
of anisotropy in the CMB on ten-degree scales would be lower than is observed
\citep{Knox98,Shaver99}.

The mass function of the first stars is very uncertain.  \citet{Kashlinsky83}
argued that the main constituents of primordial objects would be low mass
stars, with a few very massive objects (VMOs) of masses $10^3 - 10^5\,\Msun$
forming at later stages.  Other authors \citep{Carr84,Loeb97} have argued that
the lack of cooling by metals increases the Jeans mass and consequently star
formation is biased towards the production of high mass stars.  Some recent
numerical studies \citep{Omukai98,Bromm99} find that the first generation stars
are quite massive.  Supernovae at the death of high mass stars will contaminate
the primordial gas with metals, greatly increasing the ability of the gas to
cool, as well as injecting large amounts of kinetic and thermal energy into the
pregalactic medium.

The {\Ht} molecule, with a binding energy of only $4.48\,\mathrm{eV}$, is easily
dissociated and UV photons produced by the first objects could quickly destroy
all {\Ht}, even accounting for self-shielding \citep{Haiman97}. It has been
suggested that this might delay further structure formation until larger
objects form with non-linear mass scales above $10^8\,\Msun$ and virial
temperatures of $10^4\Kelvin$ where hydrogen line cooling dominates. An
alternative view is that the elevated electron abundance occurring in the
photo-ionized regions around massive stars provides ideal conditions for the
re-formation of a substantial {\Ht} fraction as these regions cool and
recombine. In the latter case we might anticipate that further cooling and
condensation would be promoted. 

In order to assess the nature of the first objects and importance of
{\Ht}-mediated cooling, many authors have simulated the formation of primordial
clouds.  The computational expense of following serveral chemical rate
equations in 3-D hydrodynamic calculations has resulted in the use of
one-dimensional models of cloud collapse by a number of authors.
\citet{Lepp84}, for example, used a spherical model for cloud collapse
and followed the chemical reactions responsible for {\Ht} formation. A similar,
high resolution (but still 1-D) calculation was undertaken by \citet{Haiman96}.
\citet{Tegmark97} used an even more computationally economical, semi-analytic
approach which runs so quickly that the collapse mass versus redshift parameter
space could be extensively explored.

One-dimensional approaches, however, neglect the complicated filamentary and
sheet-like structures ubiquitous to three-dimensional simulations of CDM
universes. With improvements in numerical techniques and computer hardware it
is now feasible to undertake realistic high resolution 3-D hydrodynamic
simulations which incorporate the relevant chemical reactions.  The full
three-dimensional approach has been taken by \citet{Anninos97}, and is taken
here. Despite the recent successes of 3-D numerical approaches, significant
challenges still remain. Adequate modelling of feedback from photio-ionizing
radiation and supernova blast waves has not yet been addressed fully
self-consistently in a numerical simulation although important first steps have
been taken (see for example \citealt{Gnedin97} and \citealt{Haiman99}). Second, the
shallow slope of the CDM spectrum on scales $\lesssim 10^8\,\Msun$ leads to a
wide range of scales collapsing nearly co-\ae vally and very long-range
correlations in the peaks of the mass fluctuation field, posing a severe
challenge to the limited dynamic range of 3-D simulations. In this paper we
present numerical techniques for modelling the collapse and condensation of gas
in dark matter haloes at high redshift with a view to reliably identifying the
sites of first star formation. In a subsequent paper we will investigate the
effect of feedback of these first stars on the pregalactic medium.

A pre-requisite for star formation in dark-matter-dominated cosmologies is that
gas becomes self-gravitating (i.e. the baryon density becomes locally greater
than the dark matter density) although this is perhaps not a sufficient
condition.  The earliest time that stars could then form is the time at which
there exist objects that have self-gravitating regions; likely in the core.
Identifying these regions in numerical simulations is the key objective of this
paper. Our goal is to model primordial objects and determine their mass and the
time at which they become self-gravitating within cosmological simulations. 
This aim distinguishes our approach from that of \citet{Bromm99} in which the
focus is to examine the internal properties of the first self-gravitating
clouds.

Structure formation is hierarchical in the CDM model: small objects collapse
first and coalesce into larger objects.  We may indentify three main stages in
the merger hierarchy of the two-component fluid as follows (cf.
\citealt{White78}).  Initially, the potential gradients resulting from the first
small dark matter fluctuations will be very much less than pressure gradients
which can be generated in the gas, leading only to small adiabatic density
perturbations in the baryonic component.  As the dark matter haloes become
larger through merging and accretion, the potential gradients will increase to
a point at which the baryonic material is compressed to the same fractional
overdensity as the dark matter. A dark matter halo is conventionally said to
have reached the Jeans mass when this occurs. As the mass of the underlying
dark matter haloes increases beyond this point the gas becomes subject to the
dominant gravitational influence of the virializing dark matter haloes. This
will result in the baryonic matter shock-heating to the effective virial
temperature of the dark matter halo and attaining a broadly similar density
profile. The subsequent behaviour of gas in a dark matter halo depends upon the
efficiency with which it can cool. In hierarchical models the ratio of the
cooling time to the dynamical time is of crucial importance
\citep{White78,Blumenthal84}.  If $t_{cool} \ll t_{dyn}$, the gas cools
efficiently and free falls in the dark matter potential well.  Conversely, if
$t_{cool} \gg t_{dyn}$, the gas is unable to condense appreciably before being
incorporated into a larger halo as the hierarchy builds, a process which, in a
smooth hierarchy, occurs on a characteristic timescale set by $t_{dyn}$.  In
this paper we demonstrate explicitly these three stages as haloes grow in the
CDM hierarchy.

Note that the Jeans mass as used in the above sense for a two component fluid
differs from its original definition as the perturbative scale separating
acoustic oscillation from instability in a self-gravitating gas. A perturbative
treatment of the two component fluid is complicated in detail (see, for
example, \citealt{Meiksin99}), but we may make a useful simple estimate of the
mass scale of a virialized  (non-linear) dark matter halo below which baryonic
matter is not significantly perturbed, by equating the velocity dispersion of
the dark  matter in the virialized halo with the specific thermal energy of the
gas: 
\begin{equation}
\label{vdisp_thermal}
\frac{1}{2} \left< v^2\right> = \frac{3\mathrm{G}M}{5R} = \frac{3kT}{2
\mu m_\mathrm{H}}.
\end{equation}
At the Jeans mass the gas has been adiabatically compressed to the same
overdensity as the dark matter, implying a baryon temperature $T=T_b
(\delta_v+1)^{2/3}$, where $T_b$ is the background baryon temperature and
$\delta_v$ is the overdensity relative to the background at the time of
collapse. The Jeans mass is then approximately 
\begin{equation}
\label{Jeans_mass}
M_J=\left(\frac{5 k T_0 (1+z)}{\mu
m_\mathrm{H}}\right)^{3/2} \frac{\sqrt{\delta_v+1}}{2\mathrm{G}H_0},
\end{equation}
where
$T_0 (1+z)^2$ is the temperature of the unperturbed gas at redshift $z$.

The outline of the paper is as follows.  We discuss the network of chemical
reactions involved in the production of molecular hydrogen in
\S~\ref{ch_chemcool}, along with the details of its implementation in our
\nbody gravitational and gasdynamic code, Hydra \citep{Couchman95}.  As an
initial test of our  implementation we compare, in \S~\ref{ch_tests}, the
results of spherically symmetric top hat collapses with an analytic model.  In
\S~\ref{ch_collapse} we determine the minimum mass at a given redshift that an
object must have to cool efficiently and collapse to high densities where star
formation may begin.  This was accomplished by performing many different
simulations of top hat perturbations with different masses and overdensities. 
A more realistic, cosmological simulation is described in section
\S~\ref{ch_cosmo}.  Here we simulate a $(25 \mathrm{kpc})^3$ volume of space
and follow the behaviour of the baryonic component in detail through the three
stages of cooling and condensation in hierarchical models as described above.

%% file: chemistry.tex
\section{Molecular hydrogen cooling and chemistry}
\label{ch_chemcool}

The {\Ht} molecule has long been known to be an important coolant for
gravitational collapse of hydrogen clouds that are cooler than $T \sim
1000\Kelvin$.  \citet{Gould63} discussed  formation processes and cooling rates
for molecular hydrogen in the interstellar medium;  \citet{Gould64} discussed
protostellar collapse through molecular hydrogen cooling.  {\Ht} is also a key
ingredient of the globular cluster formation scenario of \citet{Peebles68}. 
The hydrogen molecule is also important for structure formation in the early
universe; the first objects to form have virial temperatures of $\sim
1000\Kelvin$, where the {\Ht} molecule is the dominant coolant.   Through {\Ht}
cooling, star clusters \citep{Couchman86} or VMOs \citep{Bond84} could form and
subsequently reionize the intergalactic medium.  Molecular hydrogen is also
essential to the formation of the first generation stars \citep{Stahler86}.

In addition to hydrogen, the primordial gas contained {\Li}, {\D}, {\Het}, and
{\Hef} with abundances of $10^{-10}$, $10^{-5}$, $10^{-5}$, and $0.24$
respectively, relative to hydrogen.  Line cooling is dominated by neutral
hydrogen because the excitation rates of {\Li}, {\D}, and {\Het} are much
smaller \citep{Abel97a}.  

Molecular hydrogen is the most common molecule at the epoch at which the first
primordial stars form \citep{Lepp84}.  In the ISM molecular hydrogen formation
occurs on the surface of grains, but, since stars produce the grains, this
cannot be an important mechanism for the first objects.  There are, however,
other  chemical reactions that are able to form {\Ht}, which are discussed
below. At a redshift of 100, the fractional abundance of molecular hydrogen was
$10^{-6}$ \citep{Galli98}.   Other molecules that were  important coolants are
{\HD} and {\LiH} which had abundances $10^{-3}$ and $10^{-4}$ times lower than
that of \mbox{{\Ht}.}  The {\HD} molecule becomes the dominant coolant at
$20\Kelvin \le T \le 100\Kelvin$, and {\LiH} dominates below
$20\Kelvin$.   These two molecules will be important in the coldest and
densest gas in an object, and to get to that state the gas initially will
radiate thermal energy through {\Ht} cooling.  Molecular hydrogen cooling is
efficient enough to produce self gravitating objects, which is a necessary
condition for star formation.  Since we are interested in following the
evolution of primordial objects only to the self gravitating state where
$\rho_b \approx \rho_{dm}$ and not to higher baryonic densities, we need
include only {\Ht} cooling and can ignore {\HD} and {\LiH} cooling.

\subsection{Cooling processes}

When an {\Ht} molecule collides (usually with neutral hydrogen, the most
abundant species) and becomes rotationally or vibrationally excited, it may
radiatively de-excite which results in gas cooling, or it may collisionally
de-excite which results in no net energy loss from the gas.  {\Ht} vibrations
are excited at $6000\Kelvin$; below $3000\Kelvin$ only the rotational
levels of the molecules can be excited \citep{Puy93}. Radiative de-excitation
dominates  at the low densities relevant for primordial gas.  In this case,
{\Ht} molecules are mainly in the ground or first ($J=1$) rotational state.  
Collisional excitations are quickly followed by radiative decay.  At high
densities ($n=10^3$ to $10^4\,\mathrm{cm}^{-3}$ for $10^2\Kelvin \le T \le
10^4\Kelvin$), collisions dominate and Boltzmann populations are reached. 
These densities are achieved in the late stages of the collapse of primordial
objects. 

The cooling mechanisms at various temperatures are shown in
figure~\ref{coolcurve}.  At temperatures between $10^2\Kelvin$ and
$10^4\Kelvin$ the {\Ht} molecule is the most effective coolant. 
\citet{Hollenbach79} express the {\Ht} cooling function $\Lambda_{\Ht}$ in the
form 
\begin{equation} \Lambda_{\Ht}[n_H,T]=\frac{\Lambda_{LTE}}{1+n_{cr}/n_H},
\end{equation} 
where $\Lambda_{LTE}$ is the LTE cooling function (see \citealt{Hollenbach79} for
the full expression), $n_{cr}$ is a critical density defined by:
\begin{equation}
	\frac{n_{cr}}{n_H}=\frac{\Lambda_{LTE}}{\Lambda_{\Ht}[n_H\rightarrow 0]},
\end{equation} 
and $\Lambda_{\Ht}[n_H\rightarrow 0]$ is the low-density limit
of the cooling function.  The {\Ht} cooling function has been computed by
\citet{Martin96} for $T > 600\Kelvin$ and \citet{Forrey97} for $T <
600\Kelvin$.  \citet{Galli98} provided a fit to these two computations in
the low-density limit, which we adopt here: \begin{equation} \label{Htcool} \log\Lambda_{\Ht}[n_H
\rightarrow 0] = -103.0 + 97.59 \log T -48.05 (\log T)^2 +10.80(\log T)^3
-0.9032(\log T)^4. \end{equation} 

There is uncertainty in the {\Ht} cooling function because of the difficulty in
calculating the interaction potential at low temperatures.  Different choices
for rotational and vibrational {\Hn}-{\Ht} rate coefficients will produce
differences in $\Lambda_{\Ht}$;  figure~\ref{Htcoolcurves} illustrates {\Ht}
cooling rates computed by several authors.

At temperatures between $10^4\Kelvin$ and $5\times 10^4\Kelvin$
radiational de-excitation of {\Hn} dominates the cooling function.  We use the
expression given by \citet{Dalgarno72}:
\begin{equation} 
	\label{Hlinecool} 
	\Lambda_{Hline} \sim 7.5 \times 10^{-19}e^{-118348\Kelvin/T} 
	n^2 \fel (1-\fel).
\end{equation} 
Cooling processes such as He line cooling and Bremsstrahlung
(fig.~\ref{coolcurve}) that operate at higher temperatures may be ignored,
since the primordial objects have virial temperatures of $\sim 1000\Kelvin$.

\input{\thisdir/texfigs/coolcurves}

\input{\thisdir/texfigs/H2coolcurves}

\subsection{Molecular hydrogen formation and destruction}

\input{\thisdir/table1}

Despite a relatively small number of elements in the primordial gas, there is a
complicated network of reactions involving these elements; \citet{Galli98} list
87 reactions in their comprehensive study of the chemistry of the early
universe.  Fortunately many of these do not affect significantly the collapse
of primordial objects and the number of important reactions is much smaller. 
The {\HD} and {\LiH} molecules are not important at densities in our regime of
interest.  Helium chemistry and cooling becomes important above
$10^5\Kelvin$, which is well above the $1000\Kelvin$ virial temperatures that
primordial objects have.  Ignoring {\HD}, {\LiH}, and {\He} molecular species
reduces the number of relevant reactions considerably, and will only limit
investigations of the latest stages of baryonic condensation at low
temperatures which are not the focus of this paper. We summarize and justify in
this section the reactions which it is necessary to include in our numerical
model to accurately follow the chemistry and cooling of the first objects. 

Molecular hydrogen production proceeds via only two major mechanisms.  In the
{\Htp} channel the proton acts as a catalyst: \begin{eqnarray}  \Hp+\Hn
&\longrightarrow& \Htp + \chemical{\gamma}\nonumber\\ \Htp + \Hn 
&\longrightarrow& \Ht + \Hp. \nonumber  \end{eqnarray}  In the {\Hm} channel
the electron acts as a catalyst:   \begin{eqnarray}  \Hn+\el &\longrightarrow&
\Hm + \chemical{\gamma}\nonumber\\ \Hm + \Hn  &\longrightarrow& \Ht +
\el.\nonumber \end{eqnarray}  For redshifts higher than 200, CMB photons are
energetic enough that the reaction ${\Hm} + \gamma \longrightarrow {\Hn} +
{\el}$ neutralizes all {\Hm} ions and therefore the {\Hp} channel dominates for
$z \gtrsim 200$.  At lower redshifts, the {\Hm} channel proceeds much more
quickly than the {\Hp} channel and dominates {\Ht} production.

The destruction of the intermediaries {\Hm} and {\Htp} proceeds much more
quickly than their production.  This leads to equilibrium fractional abundances
of {\Hm} and {\Htp} (relative to total hydrogen number density) of: 
\begin{eqnarray} 
\fHm & = & \frac{\rHnelHmg \fHn \fel + \rHtelHnHm \fHt \fel}
{ \rHmHnHtel  \fHn  +  \rHmHpTHn   \fHp + \rHmelHnTel \fel +
  \rHtpHmHtHn \fHtp +  \rHmHpHtpel \fHp +
	\rHmHnTHnel \fHn} \label{Hmeqabund}\\
\fHtp & = & \frac{ \rHtHpHtpHn \fHt \fHp + \rHpHnHtpg \fHp + \rHmHpHtpel \fHm
\fHp}{ \rHtpHnHtHp + \rHtpHmHtHn \fHm + \rHtpelTHn \fel},
\end{eqnarray} 
where the rate for reaction $i$ is $k_i$. Table \ref{reactions} lists the
reactions we have considered here; where more than one source is listed for the
reaction rate, we have used the rate given by \citet{Abel97b}.

Reaction rates depend on temperature and the abundances of the reactants, all
of which  change with time, and therefore these must be known to identify the
dominant reactions.  This leads to a circular argument: the density and
temperature must be known to identify dominant reactions, but the dominant
reactions must be known to properly model the density and temperature
evolution.  One could include every chemical reaction, but this is
computationally expensive.  Instead, we used reasonable estimates for the
chemical abundances, density, and temperature to tentatively identify the
important reactions, and included these in our \nbody code, along with the
dominant cooling mechanisms; the implementation is discussed in
\S\ref{sec_coding} below.  Then we performed a spherically symmetric top hat
collapse of $10^6\,\Msun$ test object (see \S\ref{ch_tests} for
details) following the temperature and abundances
during the evolution to plot the
reaction rates versus redshift in figures~\ref{H2rates}-\ref{Hprates}.  From
these figures the unimportant reactions may be identified, and their exclusion
justified. We believe that our identifications are robust.

Figure~\ref{H2rates} shows the {\Ht} production and destruction mechanisms
along with the temperature of the collapsing object as a function of redshift.
The peak temperature occurs at $1+z=20$ when the test object
virializes.  Although this redshift was arbitrarily chosen, the behaviour of
the reactions does not change significantly for perturbations that virialize
between $50 \lesssim z \lesssim 10$, the redshifts at which the first objects
are expected to form in models with CDM-like spectra.  It is
immediately obvious from 
figure~\ref{H2rates} that the only {\Ht} production mechanism of importance  at
these redshifts is the {\Hm} channel.  The {\HTp} abundance is less than
$10^{-16}$ \citep{Galli98} so reaction \ref{HTpelHtHn} is unimportant.  Since
the {\Htp} abundance does not influence the {\Ht} production significantly,
reactions \ref{HtpHnHtHp}, \ref{HtpHmHtHn}, and \ref{HtpelTHn} may be
ignored.  

Figure~\ref{Hmrates} shows the reactions involving the {\Hm} ion.  Reactions
\ref{HnelHmg} and \ref{HmHnHtel} (the {\Hm} channel) are in equilibrium and no
other reactions proceed at comparable rates.  At $T > 10^4\Kelvin$  the
{\Hm} ion is quickly destroyed by collisions with electrons (reaction
\ref{HmelHnTel}), thus terminating {\Ht} production through this route.

At the low temperatures ($T \lesssim 5000\Kelvin$) encountered in the the
first bound objects, there is no efficient means of destroying {\Ht}.  Although
unimportant for objects with virial temperatures of $T \lesssim
1000\Kelvin$, {\Ht} dissociation mechanisms must be included for the
reaction network to be accurate at higher temperatures.  The five {\Ht}
destruction mechanisms in figure~\ref{H2rates} all have similar rates;
however at $T=5000\Kelvin$ protons begin to dissociate {\Ht} via reaction
\ref{HtHpHtpHn}.  Once the free electron abundance becomes significant at $T >
2 \times 10^4\Kelvin$, {\Ht} dissociation by electrons dominates, with the
production of two neutral hydrogen atoms and an electron (reaction
\ref{HtelTHnel}) strongly favoured over the production of a neutral hydrogen
and a negative hydrogen ion (reaction \ref{HtelHnHm}).  The dissociation of
{\Ht} by {\Hn} (reaction \ref{HtHnThHn}) is unimportant since almost all
hydrogen is ionized at $T > 2 \times 10^4\Kelvin$, at which point {\Ht} is
dissociated entirely by collisions with free electrons (reaction
\ref{HtelTHnel}).  At lower temperatures dissociation of {\Ht} by {\Hn} is
unimportant since charge exchange with {\Hp} (reaction \ref{HtHpHtpHn}) occurs
at a greater rate.  {\Ht}-{\Ht} and {\Ht}-{\Htp} collisions are rare because of
their low abundance so reactions \ref{HtHtHtTHn} and \ref{HtHtpHTpHn} are
negligible.  

Since {\Ht} production depends on the free electron abundance, reactions
involving the ionization of hydrogen must also be examined.  The neutralization
of {\Hp} by an electron (reaction \ref{HpelHng}) is in equilibrium with
ionization of {\Hn} by an electron (reaction \ref{HnelHpTel}) in
figure~\ref{Hprates}, and these two reactions control the degree of hydrogen
ionization.  

The dominant reactions thus \ref{HpelHng} through \ref{HtHpHtpHn}.  All
reactions involving helium have been disregarded, which is permissible so long
as the temperature is restricted to $T \lesssim 3 \times 10^4\Kelvin$ where
helium begins ionization and becomes an important coolant.   All molecules
other than {\Ht} (such as {\LiH} and {\HD} and those  involving {\He}) have not
been included since their abundances are negligible.

\input{\thisdir/texfigs/ratefigs}

The equations that describe the evolution of the various species can be
determined using the dominant reactions discussed above.  The total
particle abundance is \chemical{{\it n}={\it n}[H]+{\it n}[H^+]+2{\it
n}[H_2]} and the fractional abundance of species A is \chemical{{\it f_A} =
{\it n}[A]/{\it n}.}  The {\Hn}, {\Hp}, and {\Ht} fractional abundances evolve
as:
\begin{eqnarray}
 \label{Hnrate}
 \frac{1}{n}\frac{d\fHn}{dt} & = & \rHpelHng \fHp \fel - \rHnelHpTel \fHn \fel \\
 \label{Hprate}
 \frac{1}{n}\frac{d\fHp}{dt} & = & \rHnelHpTel \fHn \fel  - \rHpelHng \fHp \fel \\
 \label{Htrate}
 \frac{1}{n}\frac{d\fHt}{dt} & = & \rHmHnHtel \fHm \fHn - \rHtHpHtpHn \fHt \fHp 
											 - \rHtelTHnel \fHt \fel
\end{eqnarray}
Since the {\Hm} formation and destruction reactions are in equilibrium it is
never necessary to integrate the chemical equations to determine its
abundance.  Note that the {\Hm} equilibrium abundance (equation
\ref{Hmeqabund}) may be simplified by including only the rate coefficients for
the dominant reactions (i.e. \rHnelHmg, \rHmHnHtel, \rHmelHnTel).  

\subsection{Photoionization and Photodissociation} 

Although ionization and dissociation by photons are important in the early 
universe, by $z=100$ there are essentially no CMB photons with energies above 
the thresholds for reactions \ref{HngHpel} ($13.6\,\mathrm{eV}$) and 
\ref{HmgHnel} ($0.755\,\mathrm{eV}$).  Since {\Htp} becomes important in
molecular hydrogen production only at redshifts $z > 200$,  reactions
\ref{HtpgHnHp} and \ref{HtpgTHpel} may be neglected. The dissociation
reactions  \ref{HtgHtpel},  \ref{HtgHpHnel}, \ref{HtgHnHn}, and \ref{HtgTHn}
are inhibited because the CMB photons are much less energetic than the {\Ht}
binding energy of $4.48\,\mathrm{eV}$.  Since there are no sources of energetic photons,
all photoionization and photodissociation reactions may be ignored from $z=100$ 
until after the first objects have been formed.

\subsection{Shocks}

There are several issues concerning {\Ht} chemistry that must be addressed when
the gas shocks during collapse.  The {\Ht} molecule has a small binding energy
of $4.48\,\mathrm{eV}$, and if a shock has a velocity of $v \gtrsim \sqrt{2 \times
4.48\,\mathrm{~eV} / (2m_p)} \simeq 20\,\mathrm{km/s}$ then {\Ht} may be collisionally
dissociated \citep{Kwan77}.  The first objects to form will have masses between
approximately $10^5\,\Msun$ and $10^6\,\Msun$ and virial temperatures of $T
\approx 1000\Kelvin$.  This temperature corresponds to an average thermal velocity 
$v = \sqrt{(3kT/2)/(m/2)}$ which is 5\,km/s for {\Hn} and
3.5\,km/s for {\Ht}, well below the collisional dissociation
velocity. We have not observed any cases in which the shock velocity
approaches the dissociation velocity in our simulations of primordial
objects. 

\subsection{Incorporation of the chemical reactions into Hydra}
\label{sec_coding}

Our $N$-body gravitational and gasdynamic code, Hydra, is an adaptive mesh
implementation of the smoothed particle hydrodynamics (SPH) algorithm and the
particle-particle particle-mesh ($\mathrm{P}^3\mathrm{M}$) algorithm.  A
complete description of the code is given by \citet{Couchman95}. The chemical
state of the gas is carried by assigning each SPH particle a fractional
abundance of {\Ht} and {\Hp} as described below.

If the macroscopic changes caused by gravitational and adiabatic heating or
cooling  operate much more slowly than the chemical processes, then the two
become decoupled and may be integrated separately.  This is indeed the
case for our simulations. 
During a spherical top hat collapse, the maximum values of $T/\dot{T}$ and
$\rho/\dot{\rho}$ are on the order of $10^7$ and $10^8$ years, while
${\fHp}/\dot{\fHp}$ is approximately $10^4$ years.  

The time rate-of-change of density is computed for each particle via
$\dot{\rho}=(\rho_i - \rho_{i-1})/dt$ where $\rho_i$ and $\rho_{i-1}$ are the
density values at the current and previous time step.  The value of
$\dot{\rho}$ is assumed to be constant over the time step; while this is not
strictly true, the approximation is justified since the time steps are
typically a factor of 100 smaller than the characteristic time scale
$\rho/\dot{\rho}$.   The time rate-of-change of temperature from adiabatic
processes is computed analogously: $\dot{T}_{ad,i}=(T_{ad,i} -
T_{ad,i-1})/dt$.  These two slowly varying dynamical quantities, $\dot{\rho}$
and $\dot{T}_{ad}$, are needed to compute the rapidly-varying chemical 
evolution of the system.  The cooling functions (fig.~\ref{coolcurve}),
adiabatic processes and the chemical abundance equations (\ref{Hnrate},
\ref{Hprate} and \ref{Htrate}) are then simultaneously integrated for each
particle at every time step.  We assume that the {\Hp} and {\el} abundances are
equal, which is justified provided temperatures are constrained to $T <
10^5\Kelvin$.  The {\Hm} abundance may be computed using equation
(\ref{Hmeqabund}).  The {\Htp} abundance is never needed since it is not a
reactant in any of the dominant reactions.  The neutral hydrogen abundance is
computed via the conservation equation \chemical{{\it n}={\it n}[H]+{\it
n}[H^+]+2{\it n}[H_2]}. Of the six species present in the dominant reaction
network ({\el}, {\Hn}, {\Hm}, {\Hp}, {\Ht}, {\Htp}),  only two are independent,
and each particle therefore carries values for the {\Ht} and {\Hp} abundances. 

We use the stiff ODE integrator STIFBS from Numerical Recipes \citep{Press92}. 
This routine was also chosen for a very similar problem by \citet{Haiman96} who
experimented with other Numerical Recipes routines and found STIFBS to be the
most efficient.  They also tested the LSODAR routine of \citet{Hindmarsh83} and
found STIFBS $\approx 20\%$ faster while producing identical results.  We
compared our integration of the chemical network of ODEs to a stiff ODE
integrator in the Matlab software package and found them to agree to 1 part in
$10^4$, which is more than adequate for our purposes.

%% file: texfigs/coolcurves.tex
\begin{figure}
\epsscale{\figscale}
\plotone{\thisdir/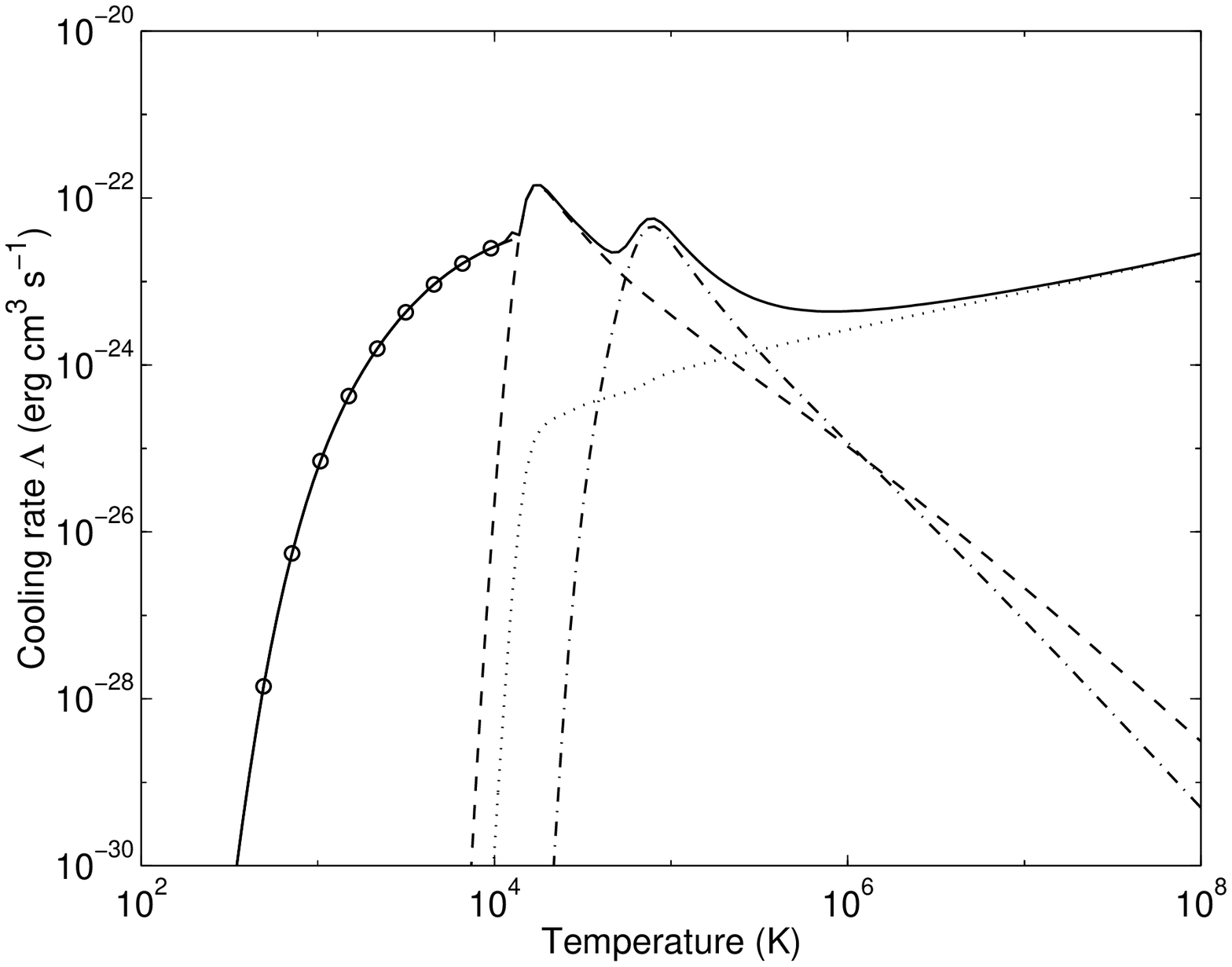}

\caption[Cooling mechanisms]{\label{coolcurve}Cooling rates for Bremsstrahlung
(dotted line), {\Hn} (dashed line) and {\He} (dash-dotted line) line cooling,
and {\Ht} (circles) cooling.  The {\el}, {\Hp}, {\Hep}, and {\Hepp} abundances were
computed assuming collisional equilibrium, and the {\Ht} fractional abundance
was $3 \times 10^{-4}$, which is typical for early objects.}

\end{figure}

%% file: texfigs/H2coolcurves.tex
\begin{figure}
\epsscale{\figscale}
\plotone{\thisdir/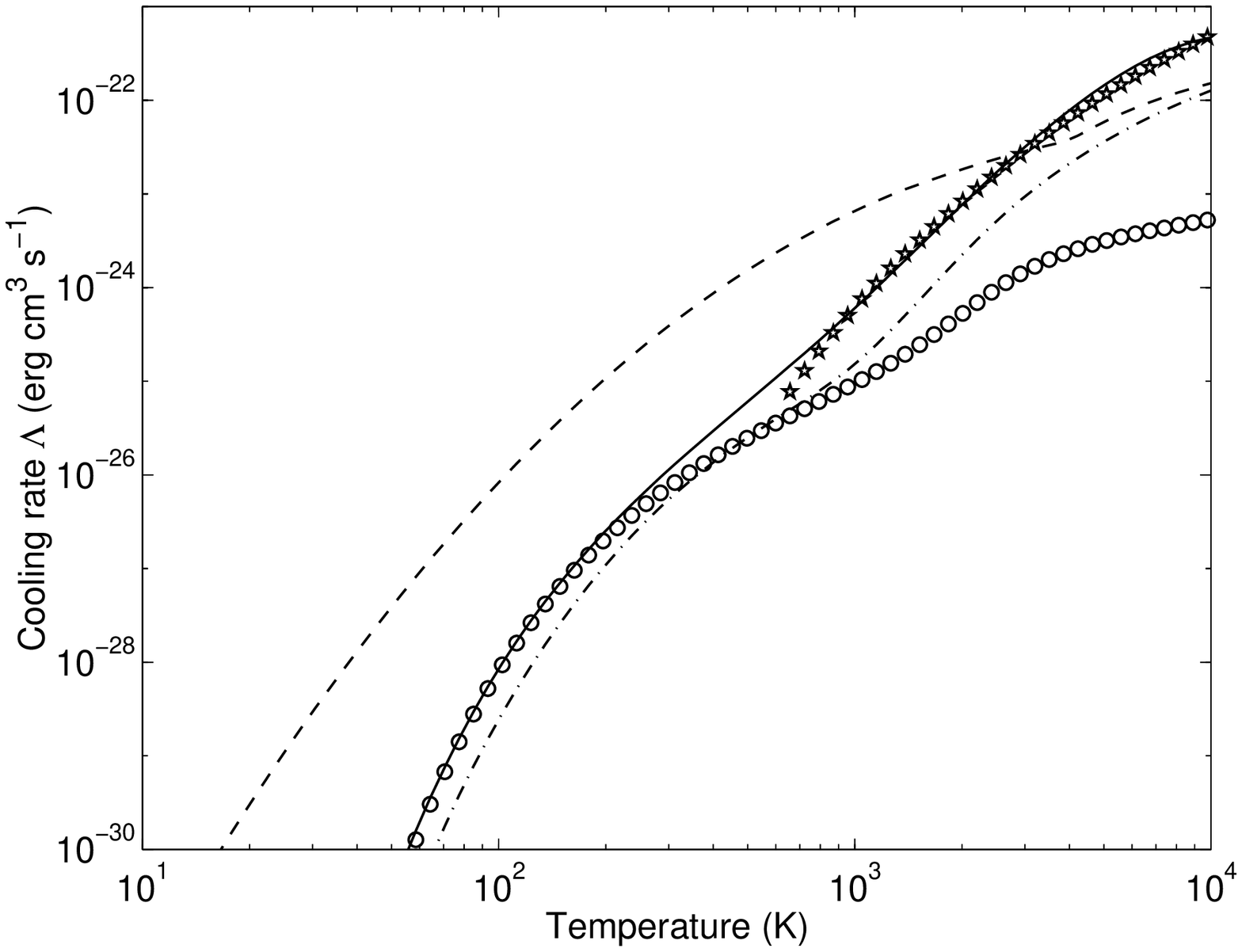}

\caption[H2 cooling curves]{\label{Htcoolcurves}{\Ht} cooling rates per {\Ht}
molecule in the low density ($n < 0.1\,\mathrm{cm}^{-3}$) limit, from:
\citealt{Galli98} (solid line), \citealt{Martin96} (stars), \citealt{Tegmark97}
(circles)., \citealt{Hollenbach79} (dash-dotted line), and \citealt{Lepp84} (dashed
line).}

\end{figure}

%% file: table1.tex
\begin{deluxetable}{rlclllr}

\tablecaption{\label{reactions} Reactions involved in the formation and destruction of molecular 
hydrogen.}
\tablehead{
\multicolumn{4}{l}{Reaction} & Rate & Reference & 
}
\startdata
\refstepcounter{tmp_eqn_count}(\arabic{tmp_eqn_count})  $\label{HpelHng}$ & 
$\Hp  + \el  $&$\longrightarrow$& $ \Hn  +  \chemical{\gamma}	$
& $1.88 \times 10^{-10}T^{-0.64}$ & \ref{Hutchins76}, \ref{Abel97a}(\ref{Ferland92})   \\
\refstepcounter{tmp_eqn_count}(\arabic{tmp_eqn_count})  $\label{HnelHpTel}$ & 
$\Hn  + \el 	$&$\longrightarrow$& $ \Hp  + 2\el 								$
& see reference & \ref{Abel97a}(\ref{Ferland92})  \\
\refstepcounter{tmp_eqn_count}(\arabic{tmp_eqn_count}) $\label{HnelHmg}$ & 
$\Hn  + \el  $&$\longrightarrow$& $  \Hm  +  \chemical{\gamma} 	$
& $1.83\times10^{-18}T^{0.88}$ & \ref{Hutchins76}, \ref{Abel97a}(\ref{Wishart79})  \\
\refstepcounter{tmp_eqn_count}(\arabic{tmp_eqn_count}) $\label{HmHnHtel}$ & 
$\Hm  + \Hn  $&$\longrightarrow$& $  \Ht  +  \el     					  $ 
&$ 1.3\times10^{-9} $& \ref{Hirasawa69}, \ref{Abel97a}(\ref{Janev87}) \\
\refstepcounter{tmp_eqn_count}(\arabic{tmp_eqn_count}) $\label{HmelHnTel}$ & 
$\Hm  + \el  $&$\longrightarrow$& $  \Hn  + 2\el 					      $
&$ 1.6\times10^{-6}(1+34/T^{5/6})\exp(-65/T^{1/3}) $&
\ref{Hirasawa69}, \ref{Abel97a}(\ref{Janev87})  \\
\refstepcounter{tmp_eqn_count}(\arabic{tmp_eqn_count}) $\label{HtelTHnel}$ & 
$\Ht  + \el 	$&$\longrightarrow$& $ 2\Hn  +  \el 								$ 
&$5.6\times10^{-11}T^{1/2}\exp(-102124/T) $& \ref{Abel97a}(\ref{Donahue91}) \\
\refstepcounter{tmp_eqn_count}(\arabic{tmp_eqn_count}) $\label{HtHpHtpHn}$ & 
$\Ht  + \Hp	$&$\longrightarrow$& $  \Htp +  \Hn 								$ 
&$ 2.6\times10^{-8}\exp(-2.12/T_4)/T^{1/2}$& \ref{Rapp62}, \ref{Abel97a}  \\
\refstepcounter{tmp_eqn_count}(\arabic{tmp_eqn_count}) $\label{HmHpTHn}$ & 
$\Hm  + \Hp  $&$\longrightarrow$& $ 2\Hn  											$ 
&$ 4\times10^{-6}/T^{1/2}$& \ref{Hill75}, \ref{Abel97a}(\ref{Dalgarno87})  \\
\refstepcounter{tmp_eqn_count}(\arabic{tmp_eqn_count}) $\label{HnHnHt}$ & 
$\Hn  + \Hn  $&$\longrightarrow$& $  \Ht   											$ 
&$ 4\times10^{-27} $ @ $T=100\Kelvin$ & \ref{Hirasawa69}  \\
\refstepcounter{tmp_eqn_count}(\arabic{tmp_eqn_count}) $\label{HtpHnHtHp}$ & 
$\Htp + \Hn	$&$\longrightarrow$& $  \Ht  +  \Hp    						$ 
&$ 6.4\times10^{-10} $& \ref{Abel97a}(\ref{Karpas79})  \\ 
\refstepcounter{tmp_eqn_count}(\arabic{tmp_eqn_count}) $\label{HtpHmHtHn}$ & 
$\Htp + \Hm  $&$\longrightarrow$& $  \Ht  +  \Hn 								$ 
&$ \leq 2.3\times10^{-7}$& \ref{Hirasawa69}, \ref{Abel97a}(\ref{Dalgarno87}) \\
\refstepcounter{tmp_eqn_count}(\arabic{tmp_eqn_count}) $\label{HTpelHtHn}$ & 
$\HTp + \el	$&$\longrightarrow$& $  \Ht  +  \Hn 								$ 
&$ 5\times10^{-9} $& \ref{Solomon71} \\
\refstepcounter{tmp_eqn_count}(\arabic{tmp_eqn_count}) $\label{HtHtHtTHn}$ & 
$\Ht  + \Ht	$&$\longrightarrow$& $  \Ht  + 2\Hn 								$ 
&$ 3\times10^{-4}\exp(-5.2/T_4)/T^{3/2}$& \ref{Hirasawa69} \\
\refstepcounter{tmp_eqn_count}(\arabic{tmp_eqn_count}) $\label{HtHmHtHnel}$ & 
$\Ht  + \Hm  $&$\longrightarrow$& $  \Ht  +  \Hn + \el 					$ 
& very small & \ref{Hirasawa69} \\
\refstepcounter{tmp_eqn_count}(\arabic{tmp_eqn_count}) $\label{HtHtpHTpHn}$ & 
$\Ht  + \Htp	$&$\longrightarrow$& $  \HTp +  \Hn 								$ 
&$ 2.1\times10^{-9}$& \ref{Hirasawa69} \\
\refstepcounter{tmp_eqn_count}(\arabic{tmp_eqn_count}) $\label{HtelHnHm}$ & 
$\Ht  + \el 	$&$\longrightarrow$& $  \Hn  +  \Hm 								$ 
&$ 2.7\times10^{-8}\exp(-4.3/T_4)/T^{3/2}$& \ref{Hirasawa69} \\
\refstepcounter{tmp_eqn_count}(\arabic{tmp_eqn_count}) $\label{HtHnThHn}$ & 
$\Ht  + \Hn	$&$\longrightarrow$& $ 3\Hn  											$ 
& see reference & \ref{Abel97a}(\ref{Dove86}) \\
\refstepcounter{tmp_eqn_count}(\arabic{tmp_eqn_count}) $\label{HtpelTHn}$ & 
$\Htp + \el  $&$\longrightarrow$& $ 2\Hn 												$ 
&$ 4.2\times10^{-8}/T^{1/2}$& \ref{Solomon71}, \ref{Abel97a}(\ref{Schneider94}) \\
\refstepcounter{tmp_eqn_count}(\arabic{tmp_eqn_count}) $\label{HpHnHtpg}$ & 
$\Hp  + \Hn  $&$\longrightarrow$& $  \Htp +  \chemical{\gamma}  $ 
&$ 1.85\times10^{-23}T^{1.8}$& \ref{Abel97a}(\ref{Shapiro87})  \\
\refstepcounter{tmp_eqn_count}(\arabic{tmp_eqn_count}) $\label{HmHpHtpel}$ & 
$\Hm  + \Hp  $&$\longrightarrow$& $  \Htp +  \el 								$ 
&$ 2.291\times10^{-0.4}$ for $T \leq 2\times10^4\Kelvin$& \ref{Abel97a}(\ref{Shapiro87}) \\
\refstepcounter{tmp_eqn_count}(\arabic{tmp_eqn_count}) $\label{HmHnTHnel}$ & 
$\Hm  + \Hn  $&$\longrightarrow$& $ 2\Hn  +  \el 								$ 
& see reference & \ref{Abel97a}(\ref{Janev87}) \\
\refstepcounter{tmp_eqn_count}(\arabic{tmp_eqn_count}) $\label{HngHpel}$ & 
$\Hn  + \chemical{\gamma} $&$\longrightarrow$& $ \Hp  + \el 		$ 
& see reference &  \ref{Abel97a}(\ref{Osterbrock74}) \\
\refstepcounter{tmp_eqn_count}(\arabic{tmp_eqn_count}) $\label{HmgHnel}$ & 
$\Hm  + \chemical{\gamma} $&$\longrightarrow$& $ \Hn  + \el 						$
&$ 0.114T_{\gamma}^{2.13}\exp(-8650/T_\gamma) $& \ref{Tegmark97}, \ref{Abel97a}(\ref{Shapiro87}) \\ 
\refstepcounter{tmp_eqn_count}(\arabic{tmp_eqn_count}) $\label{HtpgHnHp}$ & 
$\Htp + \chemical{\gamma} $&$\longrightarrow$& $ \Hn  + \Hp 						$ 
&$ 6.36 \times 10^5\exp(-71600/T_\gamma) $& \ref{Tegmark97}, \ref{Abel97a}(\ref{Stancil94}) \\ 
\refstepcounter{tmp_eqn_count}(\arabic{tmp_eqn_count}) $\label{HtpgTHpel}$ & 
$\Htp + \chemical{\gamma} $&$\longrightarrow$& $ 2\Hp + \el 						$ 
& see reference &  \ref{Abel97a}(\ref{Shapiro87})\\
\refstepcounter{tmp_eqn_count}(\arabic{tmp_eqn_count}) $\label{HtgHtpel}$ & 
$\Ht  + \chemical{\gamma} $&$\longrightarrow$& $ \Htp + \el 						$ 
& see reference &  \ref{Abel97a}\\
\refstepcounter{tmp_eqn_count}(\arabic{tmp_eqn_count}) $\label{HtgHpHnel}$ & 
$\Ht  + \chemical{\gamma} $&$\longrightarrow$& $ \Hp + \Hn + \el				$ 
& see reference &  \ref{Couchman86b} \\
\refstepcounter{tmp_eqn_count}(\arabic{tmp_eqn_count}) $\label{HtgHnHn}$ & 
$\Ht  + \chemical{\gamma} $&$\longrightarrow$& $ \Hn(1s)  + \Hn(2s,2p) 	$ 
& see reference &  \ref{Abel97a}\\
\refstepcounter{tmp_eqn_count}(\arabic{tmp_eqn_count}) $\label{HtgTHn}$ & 
$\Ht  + \chemical{\gamma} $&$\longrightarrow$& $ 2\Hn(1s) 							$ 
& see reference &  \ref{Abel97a}\\

\enddata

\setcounter{ref_count}{0}
\tablerefs{
{\refstepcounter{ref_count}\arabic{ref_count}:~\label{Abel97a}~\citealt{Abel97a}},
{\refstepcounter{ref_count}\arabic{ref_count}:~\label{Hutchins76}~\citealt{Hutchins76}},
{\refstepcounter{ref_count}\arabic{ref_count}:~\label{Hirasawa69}~\citealt{Hirasawa69}},
{\refstepcounter{ref_count}\arabic{ref_count}:~\label{Shapiro87}~\citealt{Shapiro87}},
{\refstepcounter{ref_count}\arabic{ref_count}:~\label{Karpas79}~\citealt{Karpas79}},
{\refstepcounter{ref_count}\arabic{ref_count}:~\label{Solomon71}~\citealt{Solomon71}},
{\refstepcounter{ref_count}\arabic{ref_count}:~\label{Hill75}~\citealt{Hill75}},
{\refstepcounter{ref_count}\arabic{ref_count}:~\label{Rapp62}~\citealt{Rapp62}},
{\refstepcounter{ref_count}\arabic{ref_count}:~\label{Tegmark97}~\citealt{Tegmark97}},
{\refstepcounter{ref_count}\arabic{ref_count}:~\label{Janev87}~\citealt{Janev87}},
{\refstepcounter{ref_count}\arabic{ref_count}:~\label{Ferland92}~\citealt{Ferland92}},
{\refstepcounter{ref_count}\arabic{ref_count}:~\label{Wishart79}~\citealt{Wishart79}},
{\refstepcounter{ref_count}\arabic{ref_count}:~\label{Donahue91}~\citealt{Donahue91}},
{\refstepcounter{ref_count}\arabic{ref_count}:~\label{Dove86}~\citealt{Dove86}},
{\refstepcounter{ref_count}\arabic{ref_count}:~\label{Dalgarno87}~\citealt{Dalgarno87}},
{\refstepcounter{ref_count}\arabic{ref_count}:~\label{Schneider94}~\citealt{Schneider94}},
{\refstepcounter{ref_count}\arabic{ref_count}:~\label{Osterbrock74}~\citealt{Osterbrock74}},
{\refstepcounter{ref_count}\arabic{ref_count}:~\label{Stancil94}~\citealt{Stancil94}},
{\refstepcounter{ref_count}\arabic{ref_count}:~\label{Couchman86b}~\citealt{Couchman86b}}
}

\tablecomments{References in parenthesis are the source used by
\citet{Abel97a}.  Where more than two references are given, the reaction rate
listed was obtained from the first.}

\end{deluxetable}

%% file: texfigs/ratefigs.tex
\begin{figure}
\epsscale{\figscale}
\plotone{\thisdir/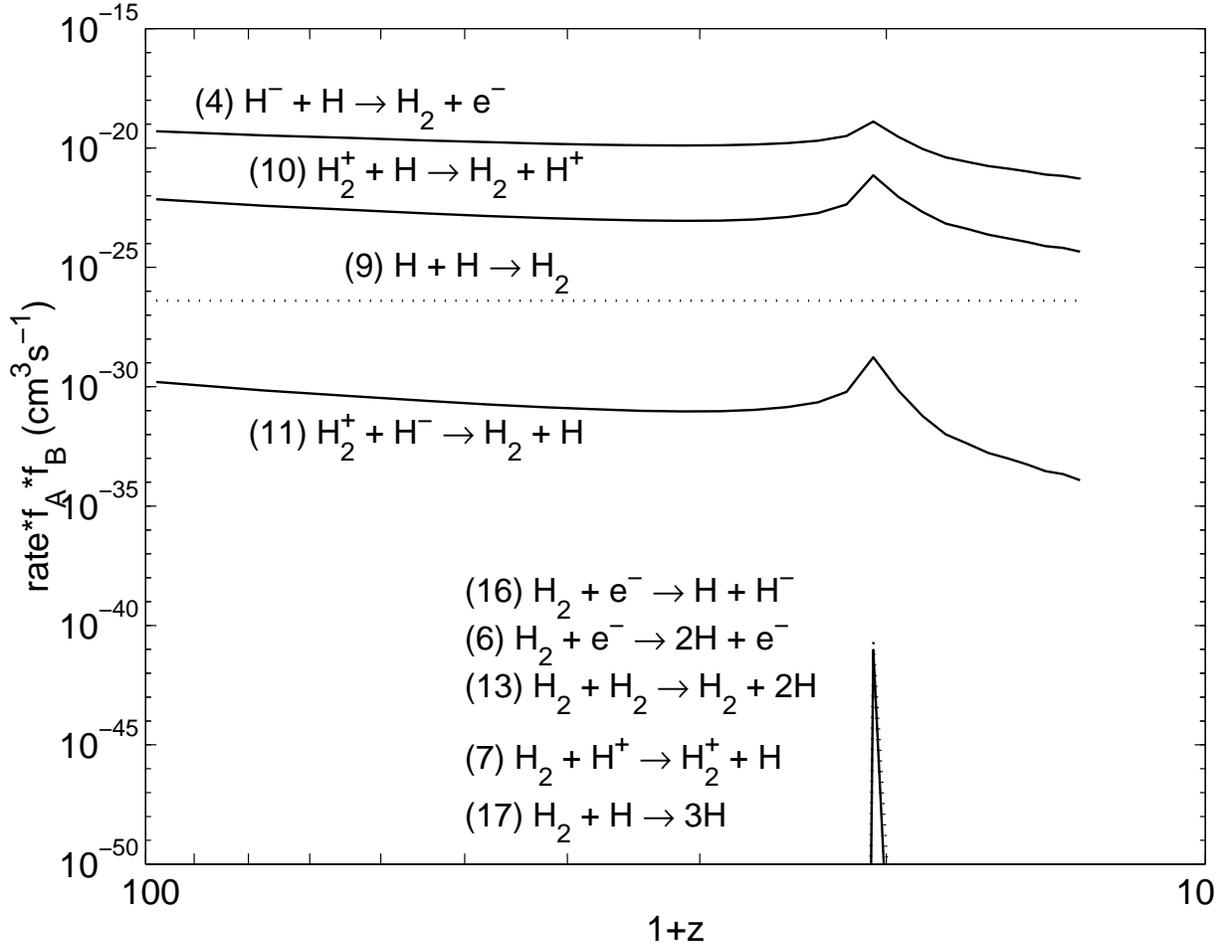}

\caption[Molecular hydrogen reactions rates]{\label{H2rates}Reaction rates for
the formation and destruction of molecular hydrogen.  Temperatures and
fractional abundances of reactants ($f_A$ and $f_B$) used to compute the rates
were taken from a top hat simulation of an early object collapsing at $1+z=20$.
{\Ht} creation via the {\Hm} channel clearly dominates.  The five {\Ht}
destruction reactions all progress at similar rates at these temperatures.  The
numbers in parentheses indicate the reaction in table \ref{reactions}.}

\end{figure}

\begin{figure}
\epsscale{\figscale}
\plotone{\thisdir/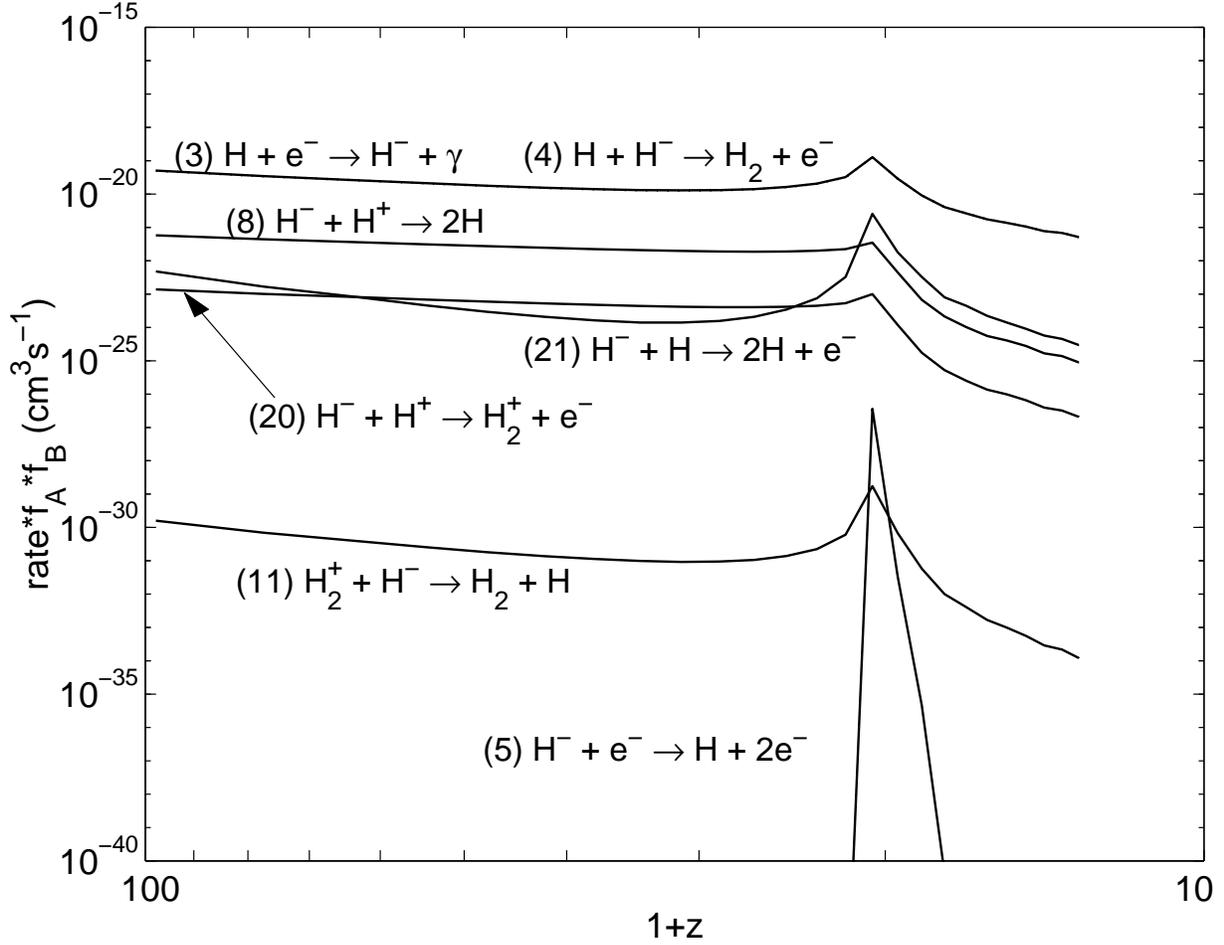}

\caption[$\mathrm{H}^-$ reaction rates]{\label{Hmrates}Reaction rates for the
formation and destruction of {\Hm}.  The two reactions in the {\Hm} channel for
molecular hydrogen production (top line) are in equilibrium.  At $T >
10^4\Kelvin$ the neutralization of {\Hm} by {\el} becomes important.  These
three reactions exclusively control the {\Hm} abundance. The
numbers in parentheses indicate the reaction in table \ref{reactions}.}

\end{figure}

\begin{figure}
\epsscale{\figscale}
\plotone{\thisdir/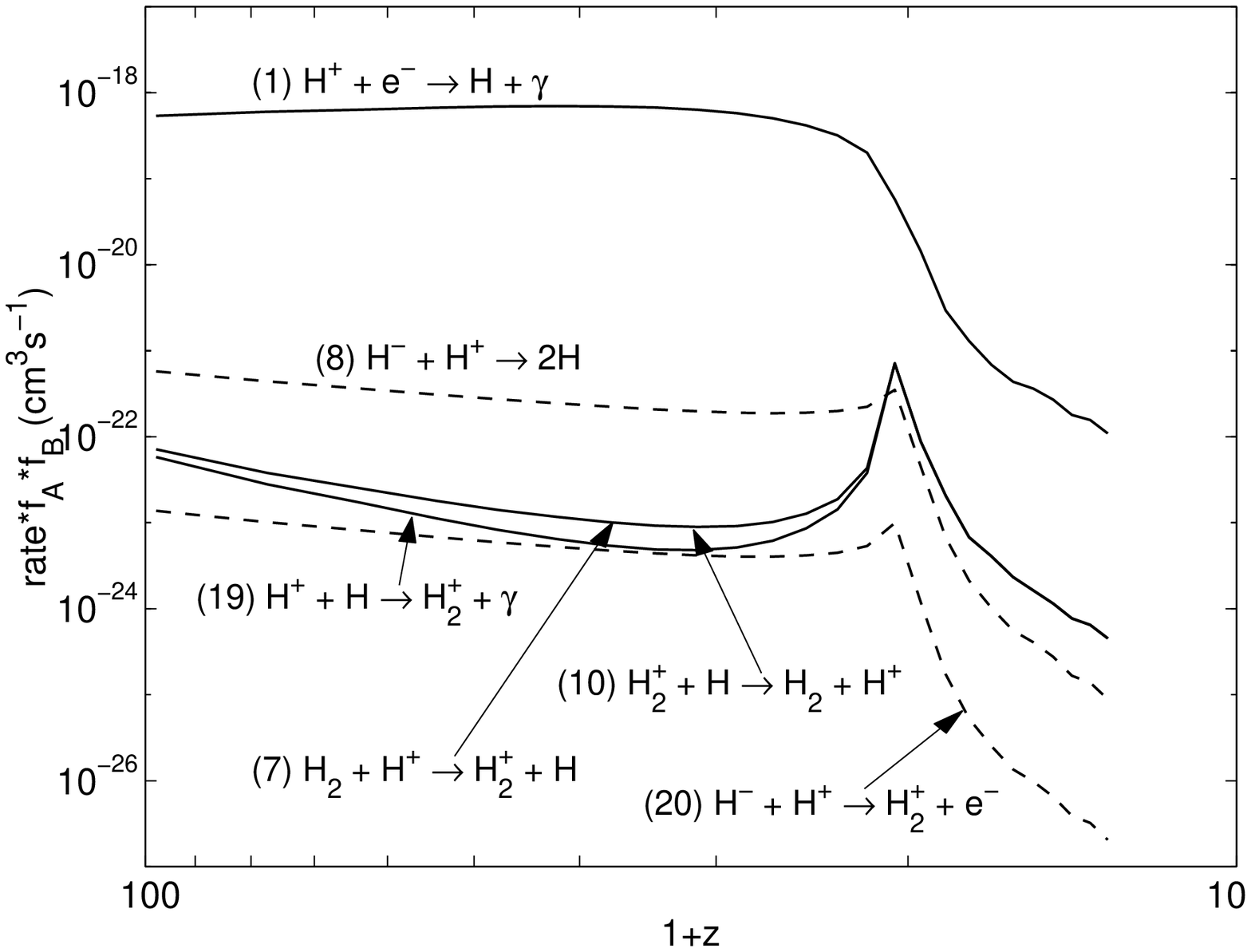}

\caption[$\mathrm{H}^+$ reactions rates]{\label{Hprates}Reaction rates
involving ionized hydrogen. Radiative recombination controls the {\Hp}
abundance.  The numbers in parentheses indicate the reaction in table
\ref{reactions}.}

\end{figure}

%% file: numerics.tex
\section{Tests of the code}
\label{ch_tests}

\subsection{Comparison with analytic models} 

In this section we present two tests: i) top hat evolution from quiet
initial conditions which proceeds to a singular collapse; and ii) a
top hat collapse in which perturbations within the top hat overdensity
lead to virialization at roughly half the maximum (turnaround) radius. 

The top hat model has two free parameters, the initial mass $M_{TH}$
and the initial overdensity which determines the redshift at which the analytic
top hat collapses to a singularity.  Using an ODE integrator in the Matlab
software package, we numerically integrated the top hat density evolution along
with relevant chemical reactions and cooling functions to analytically
determine the chemical and temperature evolution of spherical perturbations.

For a simple initial test of the chemical network in Hydra we simulate
a spherically symmetric perturbation of 
uniform density and compare that to the analytic top hat evolution.   
Particles are placed into the simulation box on a regular lattice, and a
spherical region enclosing $M_{TH}$ is cut out and compressed by an amount that
causes the singularity to be reached at the desired redshift.  Although placing
particles on a regular lattice is unphysical (because it leads to a
singularity), the absence of noise in the 
particle distribution allows an exact comparison between the code and the
analytic solution.  The particles within the top hat are assigned
velocities according to the time derivative of the cycloid equations. The gas
particles are given an increase in temperature corresponding to the adiabatic
compression.

Figure~\ref{teg_compare} shows the dark matter, baryon, temperature, {\Hm}, and
{\Ht} evolution computed using the analytic top hat model and our \nbody code
for a top hat perturbation of total mass $10^6\,\Msun$.  The simulation was
terminated prior to $z=20$ since there is no point in following the evolution
through the singularity.  This is however more than sufficient to test the
accuracy of the chemistry in the \nbody code.  We performed many other
simulations using various 
masses and virialization redshifts, and there was always excellent agreement
with the semi-analytic model up to the singularity.

\input{\thisdir/texfigs/teg_compare}

In physical reality the singularity is avoided since random
perturbations in the density 
field will cause non-radial motions to develop, and instead of collapsing to a
point the object will virialize.  The analytic evolution for the top hat
density and temperature therefore must be modified to approximate the virial
behaviour.  \citet{Tegmark97} modeled the formation of primordial objects with
a semi-analytic approach.  They assumed that the density followed the top hat
solution until the temperature reached the virial value, after which the
density remained constant at  $18\pi^2$ times the background density at the
nominal collapse redshift.  If the overdensity reached the virial value before
$T>T_{vir}$, they held the density constant and assumed that the gas would be
shock heated to the virial temperature.  This density function, the cooling
rate, and chemical reactions were integrated to follow the temperature and
chemical abundances in the top hat.

Many cosmological simulations use a regular cubic lattice for the initial
particle distribution, onto which is imposed a power spectrum of fluctuations 
according to the cosmology.  For a simple top hat model, the only fluctuation
imposed is a simple compression of a spherical region of space.  A cubic grid
is inappropriate in this case since collisionless particles would collapse to a
singularity and oscillate about the collapse centre.  
We introduce 
noise  into the particle distribution in order to generate
substructure which will lead to virialization.

To make a fair comparison between the semi-analytic top hat model and the
simulations, the initial conditions must be set up to mimic the semi-analytic
behaviour as closely as possible.  This however involves two antagonistic
criteria.  Very smooth initial conditions will not produce as much substructure
and will follow the semi-analytic top hat model closely up to the virialization
redshift.  However, the smooth initial conditions do not allow the dark matter
to virialize well: much of the dark matter rebounds outward to a large
distance while also producing a very dense, small virialized region
which has an 
overdensity well above the semi-analytic value of $\delta \simeq 200$.  Adding more
noise to the initial conditions promotes the formation of substructure which
results in a larger portion of the dark matter contained in the virialized
region and a smaller central overdensity, but the increased substructure causes
the temperature and density to rise slightly above the semi-analytic value
prior to virialization.

The amount of substructure that forms before the dark matter virializes is
governed by the noise in the initial particle positions and the time elapsed
between the start of the simulation and virialization.  Without altering the
initial particle positions, substructure in the dark matter may be
quantitatively increased by simply starting the simulation at an earlier
redshift.  One must now decide upon an appropriate amount of substructure to
use.  We have chosen to use random initial conditions and start the simulations
at a redshift that is 20 expansion times prior to virialization (eg. $z_i =
400$ for $z_v = 20$; these values are appropriate for a top hat containing
roughly $10^4$ particles).  This arrangement produces a virialized region with 
approximately $80\%$ of the dark matter within the analytic virial radius, and
a mean overdensity within the half-mass radius of approximately 200.  This
creates post-virialization conditions that are similar to the semi-analytic
model, at the expense of pre-virialization agreement.  Baryons will fall into
the shallow potential wells provided by the substructure present in the dark
matter, slightly elevating the density, temperature, and {\Ht} abundance.  The
amount of disagreement with the semi analytic model prior to virialization is
small and has negligible impact upon the post-virialization behaviour.  
Requiring the presence of substructure is also attractive since this reflects
more  closely collapses in a hierarchical CDM cosmology --- they are never
monolithic, as they are in the analytic top hat model.

We performed several top hat runs and compared with the semi-analytic
predictions. We find that after virialization, for objects that remain pressure
supported, the  simulations agree well with the semi-analytic model because the
baryon overdensity remains near the virial value.  For objects that are able to
cool and  become self gravitating, the baryon overdensities become much higher
than the virial  value, and the semi-analytic model does not agree well with
the simulations.   However, if the post-virialization baryon density in the
semi-analytic model was raised to the value observed in the simulations, the
chemical abundances and temperature agreed very well.

\subsection{Sensitivity to initial chemical abundances}
\label{init_abund}

The {\Ht} abundance in the early Universe has been computed by many authors and
varies considerably: $\approx 10^{-4}$ \citep{Tegmark97,Haiman96}, $5\times
10^{-7}$ \citep{Lepp83}, $7.5\times10^{-6}$ \citep{Puy93}.  The simulations
performed here use the {\Ht} abundance given by \citet{Anninos96}:
\begin{equation}
{\fHt}=2\times 10^{-20}\frac{ {\fHn}\Omega_0^{3/2}}{h\Omega_B}(1+z_0)^{5.1}
\end{equation}
where $z_0$ is the largest redshift at which {\Ht} can be efficiently formed
through the {\Htp} channel.  \citet{Peebles93} found the residual ionization
after recombination to be: 
\begin{equation}
{\fHp}=1.2\times10^{-5}\Omega^{1/2}/(h\Omega_B).
\end{equation}

To determine the sensitivity that the initial {\Ht} abundance has on the
collapse, two \nbody simulations of top hat perturbations were run with 
initial {\Ht} abundances of $2\times10^{-6}$ and $\sim 10^{-4}$. 
Figure~\ref{init_cond_compare} shows the evolution of the baryonic and dark
matter densities, the temperature, and the {\Ht} and {\Hp} abundances for the
central region of a top hat with $M=10^6\,\Msun$ that virializes at
$1+z_v=20$.  The {\Ht} abundance for $z \le z_v$ is not sufficient in either
run to alter the gas temperature and hence the gas evolves adiabatically: it
cools with the expansion up to the turnaround redshift $z_t=30$ and
subsequently heats as the perturbation begins to contract.  After turnaround
the increase in baryonic density accelerates the molecular hydrogen production
and the difference in the {\Ht} abundances between the two runs quickly
disappears.  Both runs reach the same final value of ${\fHt}=10^{-3}$ shortly
after virialization.  Another run was done using an initial {\Ht} abundance of
$10^{-8}$ which shortly stabilized at ${\fHt}=5\times10^{-6}$.   Prior to
virialization, there are no reactions that efficiently destroy {\Ht}, and the
factor limiting the {\Ht} abundance is the number of free electrons available
to catalyze the {\Hm} channel.  The temperature, density, and chemical
evolution after virialization are thus insensitive to the initial molecular
hydrogen abundance.  

The molecular hydrogen abundance reaches a plateau before virialization of $6
\times 10^{-6}$ which disagrees with the simulations of \citet{Haiman96} and
\citet{Tegmark97} who found the plateau values to be $1.6\times10^{-4}$ and
$10^{-4}$ respectively.  These authors  have used a rate for the
photodissociation of {\Htp} (important at $z > 200$) that is too low, which
consequently caused a higher {\Ht} abundance, as noted by \citet{Abel97b}.  Our
{\Ht} abundance agrees with the newer calculations of \citet{Galli98} and
\citet{Abel97b}.  

Regardless of the redshift at which the simulations are started, the chemical
abundances are initialized at $z=100$ and from then on the chemical reactions
and cooling processes are followed.   Computing the abundance of {\Ht} at
earlier redshifts necessitates inclusion of more chemical reaction which would
slow the code, only to produce an {\Ht} abundance at $z=100$ which is already
known (see e.g. \citealt{Galli98}). At this redshift, the top hat simulations
are still quite smooth and the baryonic density and temperature are not
significantly different from the background values.  Some error in the initial
abundances is undoubtedly committed by neglecting to compute the chemistry at
redshifts earlier than 100, but as shown above, variations in the initial {\Ht}
abundance do not alter the post-virialization behaviour.

\input{\thisdir/texfigs/init_abund_compare}

\subsection{Softening issues}
\label{soft_section}

In order to avoid excessively large forces between two particles in close
proximity, the gravitational attraction is ``softened'' so that the
gravitational force decreases at small distances.  The distance at which the
softened gravity diverges from the Newtonian value is set by the softening
length, $s$.  This parameter must be carefully tuned to the problem, otherwise
the results will be tainted by unphysical numerical effects.  If $s$ is too
small then two body collisions become apparent; if $s$ is too large then the
softening may prevent a group of particles from reaching a density high enough
to correctly model a self-gravitating region --- a necessary condition for star
formation.    SPH forces are computed by using a smoothing kernel to average
over the approximately 32 nearest neighbours; the SPH smoothing length ($h$) is
chosen adaptively such that it encloses that number of neighbours. In very
dense regions,  however, it is necessary to constrain the SPH smoothing length
to a minimum value (one half of the gravitational softening length) to balance
the gravitational and hydrodynamic resolutions, otherwise unphysical clustering
of the baryon particles can occur \citep{Thacker99}. The minimum smoothing
length determines the maximum density that  the gas particles may reach. 
Efficient cooling leads to dense clustering of the baryons relative to the dark
matter particles, and a mismatch in the ideal $s$ and $h_{min}$.  We have
performed several tests to quantify the effects that different softening and
smoothing lengths produce.

The overdensity of an object in which gas pressure is initially unimportant is,
at virialization, $1+\delta=18\pi^2$ in an $\Omega=1$ universe \citep{Gunn72}. 
The  mean separation between particles in the perturbation is then
$d=(18\pi^2)^{-1/3} d_{bg}$ where $d_{bg}$ is the mean interparticle separation
of the uniform background matter.  For a simulation box volume of $V=L^3$
containing $N=n^3$ particles, $d_{bg}=L/n$ and $d \simeq 0.2 L/n$.   The
softening length should enclose at least roughly 10 particles to minimize
spurious two body effects, which gives $s = 2 d \simeq 0.4 L/n$.  This may be
taken as an estimate of the minimum softening length, since the central region
of a virialized object has an overdensity well in excess of $\delta \simeq
200$.  As an extreme, if $\delta=10^5$, then $s \simeq 0.04 L/n$. 

The smoothing length has a much greater effect on the baryonic density than on
the dark matter density for cases in which the baryons can cool efficiently. 
This  is illustrated in figure~\ref{soft_density_collapsed} which shows the
baryonic and dark matter density profiles for top hat simulations using various
smoothing lengths (table \ref{softtab}, runs 1 to 4).   In these runs, enough
{\Ht} is formed that the gas can cool efficiently and free fall into the dark
matter potential well.  If $s$ is too large, then the baryonic density $\rho_b$
remains lower than the dark matter density $\rho_{DM}$.  The smoothing length
has the undesirable ability to determine if the gas can become self gravitating
(i.e. $\rho_b \ge \rho_{DM}$).  This problem does not exist for hot, pressure
supported objects --- in these, even an extremely small softening will not
cause the object to become self-gravitating
(figure~\ref{soft_density_supported}).

\begin{deluxetable}{cccccc}
\tablewidth{0pt}
\tablecaption{\label{softtab}Simulation parameters for softening and resolution
tests}

\tablehead{
run  & mass & box size & softening &  $z_v$ &   N\\
	   & ($10^6\,\Msun$) &  (kpc)    &  (pc)  &    \\
}

\startdata 
1 & 1.03 & 20.8 & \phantom{0}3.5 & 20 & $2\times 32^3$\\
2 & 1.03 & 20.8 & \phantom{0}7.0 & 20 & $2\times 32^3$\\
3 & 1.03 & 20.8 &           14.0 & 20 & $2\times 32^3$\\
4 & 1.03 & 20.8 &           28.0 & 20 & $2\times 32^3$\\
5 & 0.17 & 11.4 & \phantom{0}1.9 & 20 & $2\times 32^3$\\
6 & 0.17 & 11.4 & \phantom{0}3.7 & 20 & $2\times 32^3$\\
7 & 0.17 & 11.4 & \phantom{0}7.5 & 20 & $2\times 32^3$\\
8 & 0.17 & 11.4 &           14.7 & 20 & $2\times 32^3$\\
\enddata
\end{deluxetable}

Since we are concerned with finding the minimum mass of an object that can cool
efficiently and become self-gravitating, the situation above where the
softening has such a strong influence  on the baryonic density is
unsatisfactory.  A collapse criterion sensitive to numerical ({\it not}
physical) parameters is not robust.  In a pressure-supported object, the number
of particles within the softening radius is  approximately 32, while in a self
gravitating object the number of particles will be magnitudes larger. 
Figures~\ref{soft_temperature_collapsed} and \ref{soft_temperature_supported}
show the mass and temperature profiles for an object that has cooled and an
object that is pressure supported.  The number of particles (and also mass and
temperature) within the softening radius of an object that has cooled stays
fairly constant regardless of the size of the softening radius.  As an
alternative to the comparison between $\rho_b$ and $\rho_{DM}$ comparison, we
consider an object to be have collapsed to a self gravitating state if there
are more than one thousand particles within the softening radius at a
temperature of just over $100\Kelvin$.  In practice it is quite easy to
distinguish between objects that have become self gravitating and those that
are pressure supported since the former display rapid and obvious changes in
their density and temperature profiles.

\input{\thisdir/texfigs/softfigs}

\subsection{Particle resolution}
\label{resolution}

Increasing the number of particles in an \nbody simulation provides better
resolution, at the expense of computational speed.  Since we intended to
perform many simulations of top hat perturbations of different masses and
virialization redshifts in order to determine the mass required for an object
to become self gravitating at a certain redshift, it was desirable to find the
minimum number of particles that would provide acceptable resolution. This
minimum number will also provide a pointer to the particle mass resolution
necessary for robust simulations of the first objects in a cosmological
setting. 

We performed several simulations using various particle resolutions, and found
that the central density in $N=2\times 16^3$ runs was lower than that in
$N=2\times 32^3$ runs.  Some variation with resolution has to be expected since
the central regions in the lower resolution runs are marginally resolved.  In
objects where the central gas cools quickly, within the softening radius there
are about 50 dark matter and 500 gas particles in the $2\times16^3$
simulations. The $2 \times 32^3$ runs have well defined cores containing
approximately 600 dark matter and 6000 gas particles within the softening
radius.  Throughout the paper we will use the term ``core'' to describe the
regions in self gravitating objects that have constant (with respect to
distance to centre) high density and low temperature.

The $N=2 \times 32^3$ runs in which the gas is able to cool efficiently have
$N_{gas}(r)$ (cumulative number of gas particles) profiles that are flat, which
indicates that almost all of the gas particles initially contained in the top
hat perturbation lie within the softening radius.  In this case the resolution
is limited by the softening length and not the number of particles, but for the
$N=2 \times 16^3$ run the situation is reversed.  This demonstrates that the
minimum limit of acceptable particle resolution has been reached.  We performed
one top hat simulation with $N=2 \times 64^3$ and found the results agreed with
the $N=2\times 32^3$ run and are confident therefore that $2 \times 32^3$
particles provide adequate resolution.

The dark matter alone drives the gas dynamics until the baryonic material
becomes self gravitating.  The differences in the distribution in the baryonic
matter between the $N=2\times16^3$ and $N=2\times32^3$ simulations is primarily
caused by the differences in the dark matter distribution.  The mass profiles
in the $N=2\times 32^3$ runs were close to $M(r) \propto r^{3}$ as expected for
a perturbation of uniform density, but in the $N=2\times 16^3$ runs there
simply are not enough particles to accurately model the smooth density of the
top hat model, and the mass profiles deviate from being proportional to volume.

%% file: texfigs/teg_compare.tex
\begin{figure}
\epsscale{\figscale}
\plotone{\thisdir/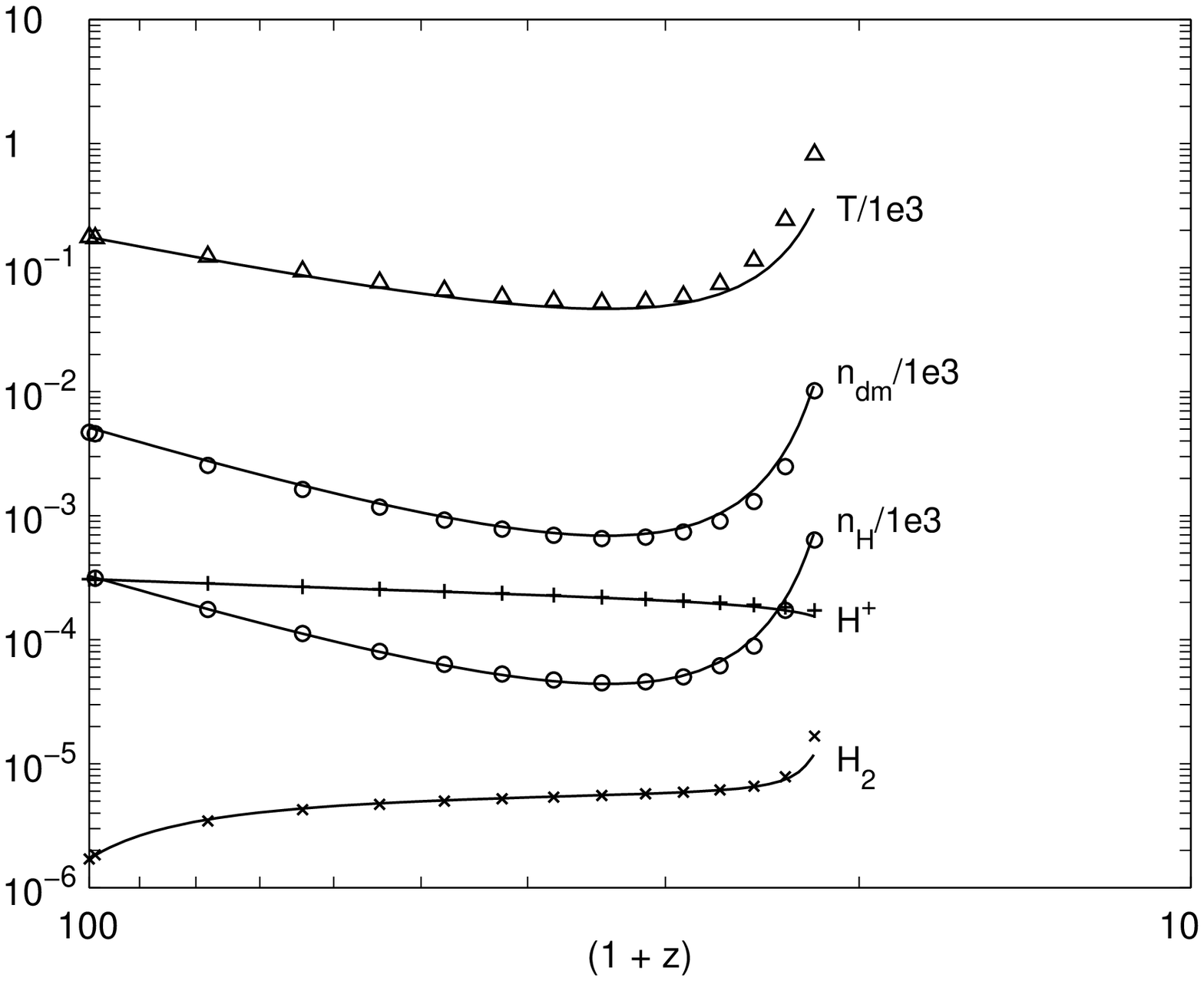}

\caption[Comparison between Hydra and analytic top hat]
{\label{teg_compare}Comparison of temperatures, baryonic and dark matter
densities, {\Hp}, and {\Ht} fractions given by an analytic top hat model
(solid lines) and our \nbody code, Hydra (symbols) for a
perturbation of total mass $10^6\,\Msun$ that virializes at $1+z=20$.}

\end{figure}

%% file: texfigs/init_abund_compare.tex
\begin{figure}
\epsscale{\figscale}
\plotone{\thisdir/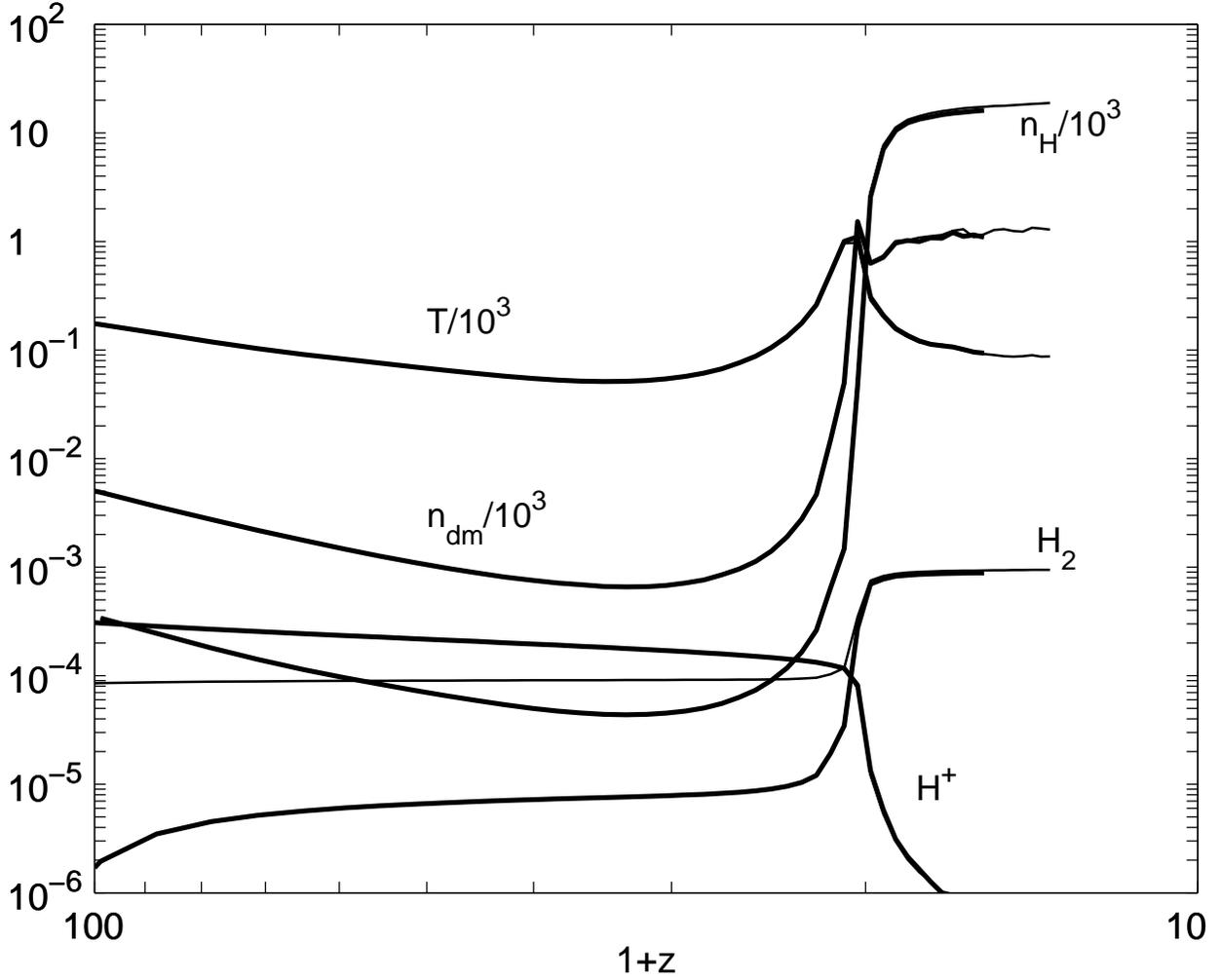}

\caption[Effect of different initial molecular hydrogen abundances]
{\label{init_cond_compare}A comparison of two runs having initial molecular
hydrogen abundances of $2\times10^{-6}$ (thick lines) and $\sim 10^{-4}$ (thin
lines).  The final physical and chemical states of the object after
virialization ($z=20$) are insensitive to the initial {\Ht} abundance.}

\end{figure}

%% file: texfigs/softfigs.tex
\begin{figure}
\epsscale{\figscale}
\plotone{\thisdir/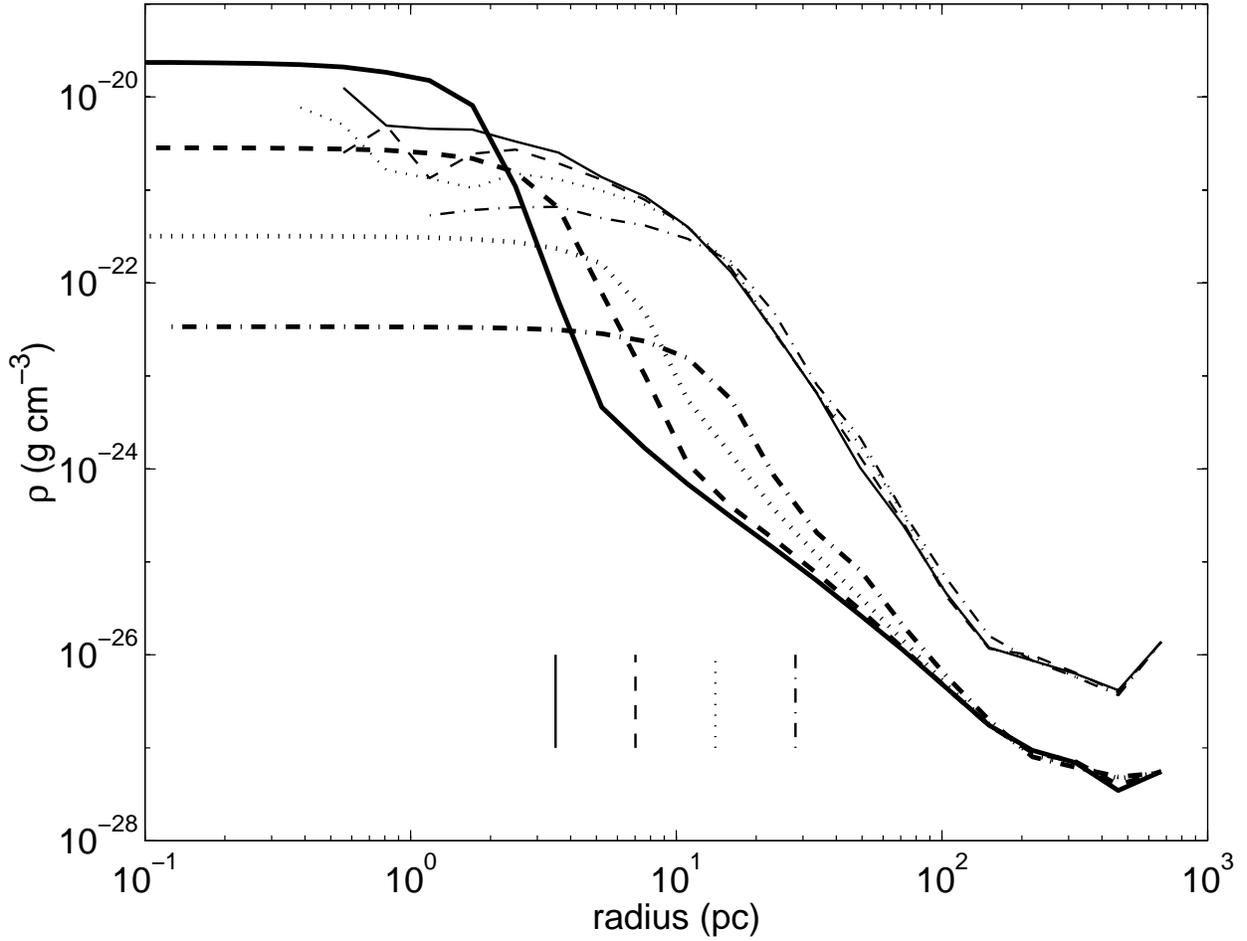}

\caption[Effect of softening on density, self gravitating object]
{\label{soft_density_collapsed}Density profiles of an object that can cool
efficiently (runs 1 to 4) for various softening lengths.  Shown are the dark
matter density (thin lines --- computed dividing the mass in a radial shell by
its volume), and the baryon density (thick lines --- computed by averaging the
SPH density of all particles within a radial shell).  The vertical lines show
the softening lengths used for each simulation.  The runs with the two smallest
softening values have self-gravitating cores ($\rho_b > \rho_{DM}$). 
Increasing the softening further causes the baryonic density to drop below that
of the dark matter density.}

\end{figure}

\begin{figure}
\epsscale{\figscale}
\plotone{\thisdir/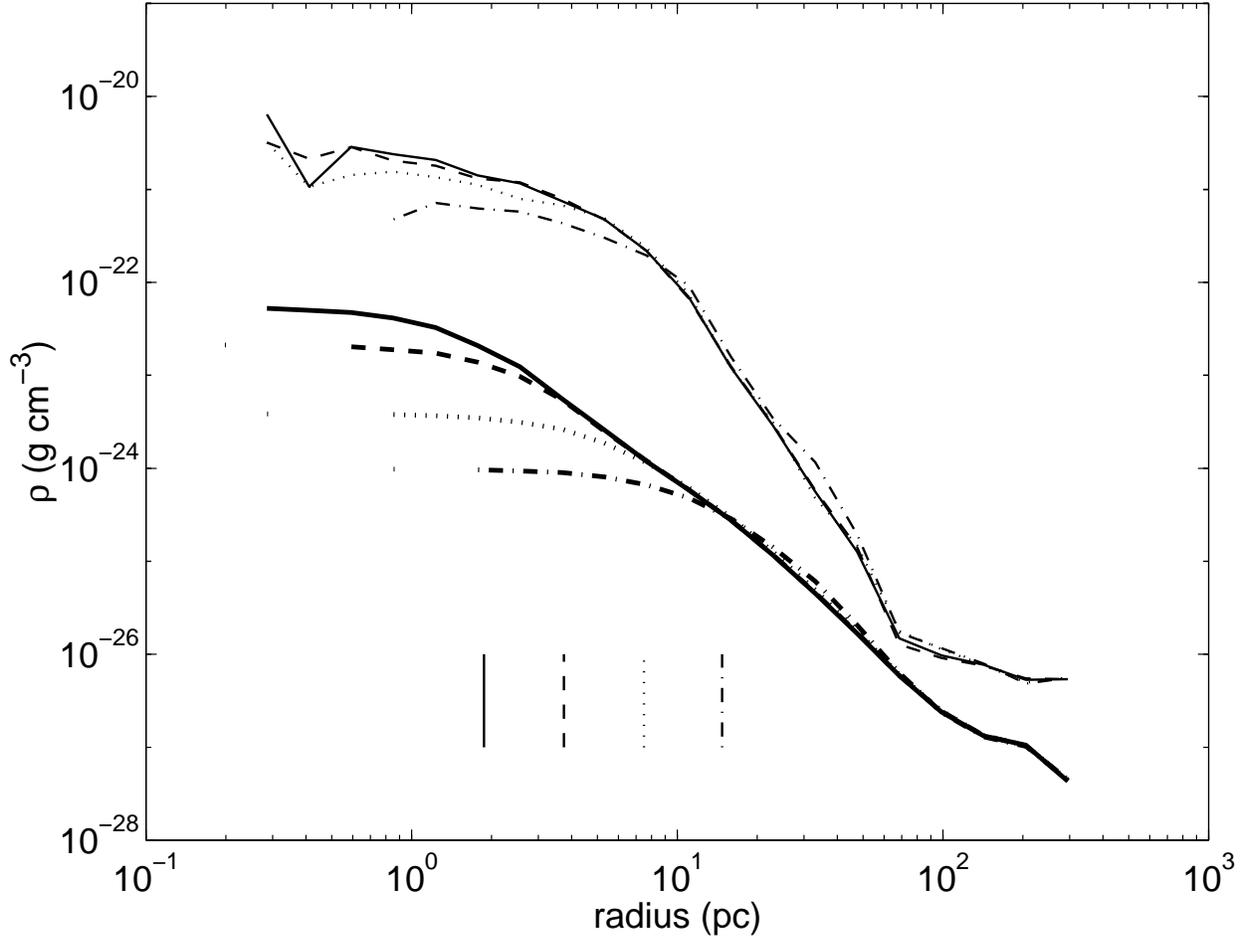}

\caption[Effect of softening on density, pressure supported lump]
{\label{soft_density_supported} Density profiles of a pressure supported object
(runs 5 to 8) for various softening lengths. In the pressure supported case,
softening does not determine if the central region becomes self-gravitating
($\rho_b > \rho_{DM}$).  As in figure \ref{soft_density_collapsed}, thick lines
are baryon densities and the thin lines are dark matter densities.  The
vertical lines show the softening lengths used for each simulation.  Densities
beyond the softening length are convergent with densities computed using a
smaller softening length.}

\end{figure}

\begin{figure}
\epsscale{\figscale}
\plotone{\thisdir/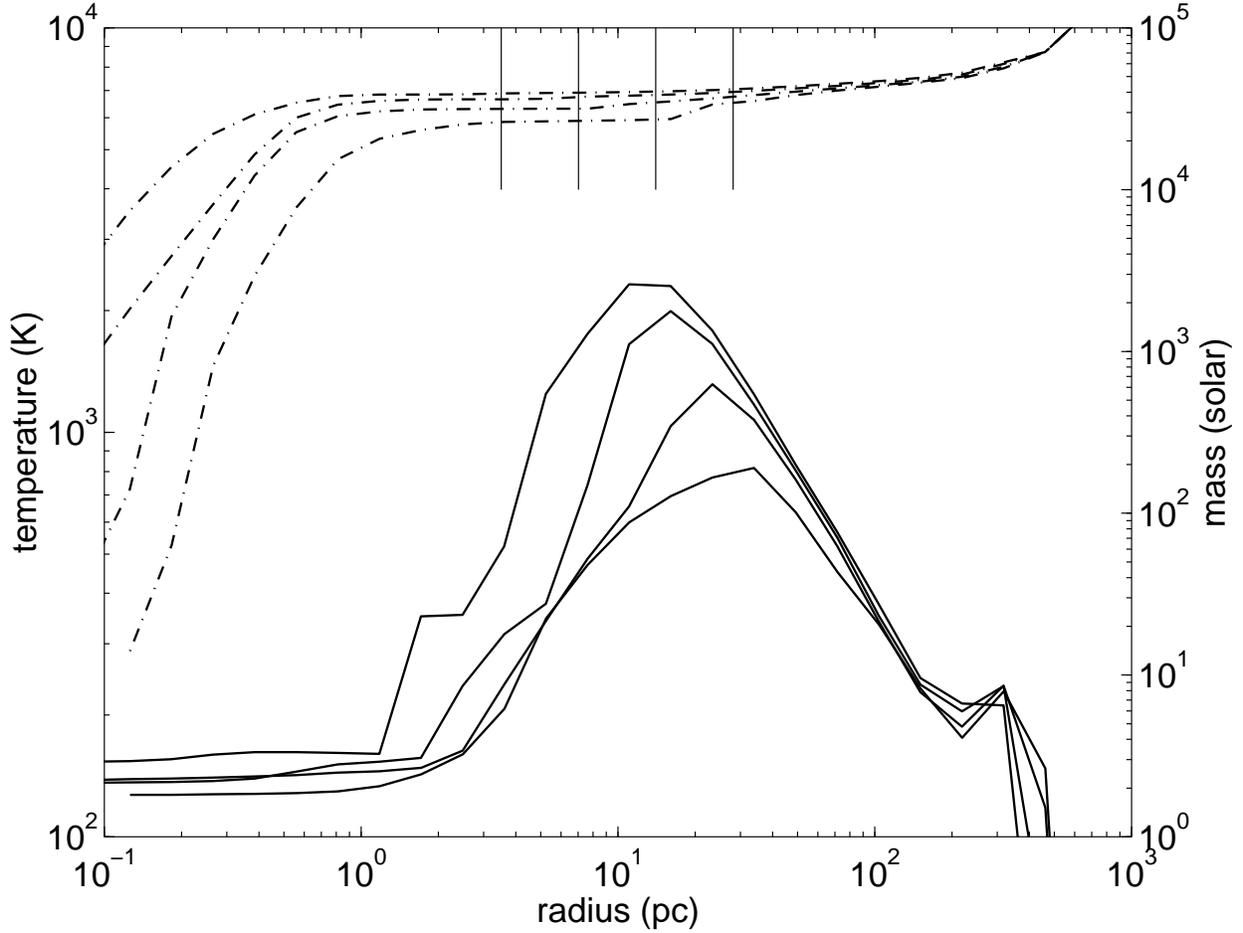}

\caption[Effect of softening on temperature, self gravitating object]
{\label{soft_temperature_collapsed}Radial profiles of gas temperature (solid
lines) and gas mass (broken lines) for various smoothing lengths (runs 1 to
4).  The object contains about 9000 particles, has a total mass of $10^6\,\Msun$
and is able to cool efficiently.  As the smoothing length (vertical lines) is
increased the peak temperature decreases, but the temperature of the baryons in
the core of the object ($r \lesssim 1 \mathrm{pc}$) is not significantly
influenced by the smoothing length.  There are approximately 3000 gas particles
within the smoothing length regardless of its value, and they have cooled to a
temperature where {\Ht} is no longer an effective coolant.}

\end{figure}

\begin{figure}
\epsscale{\figscale}
\plotone{\thisdir/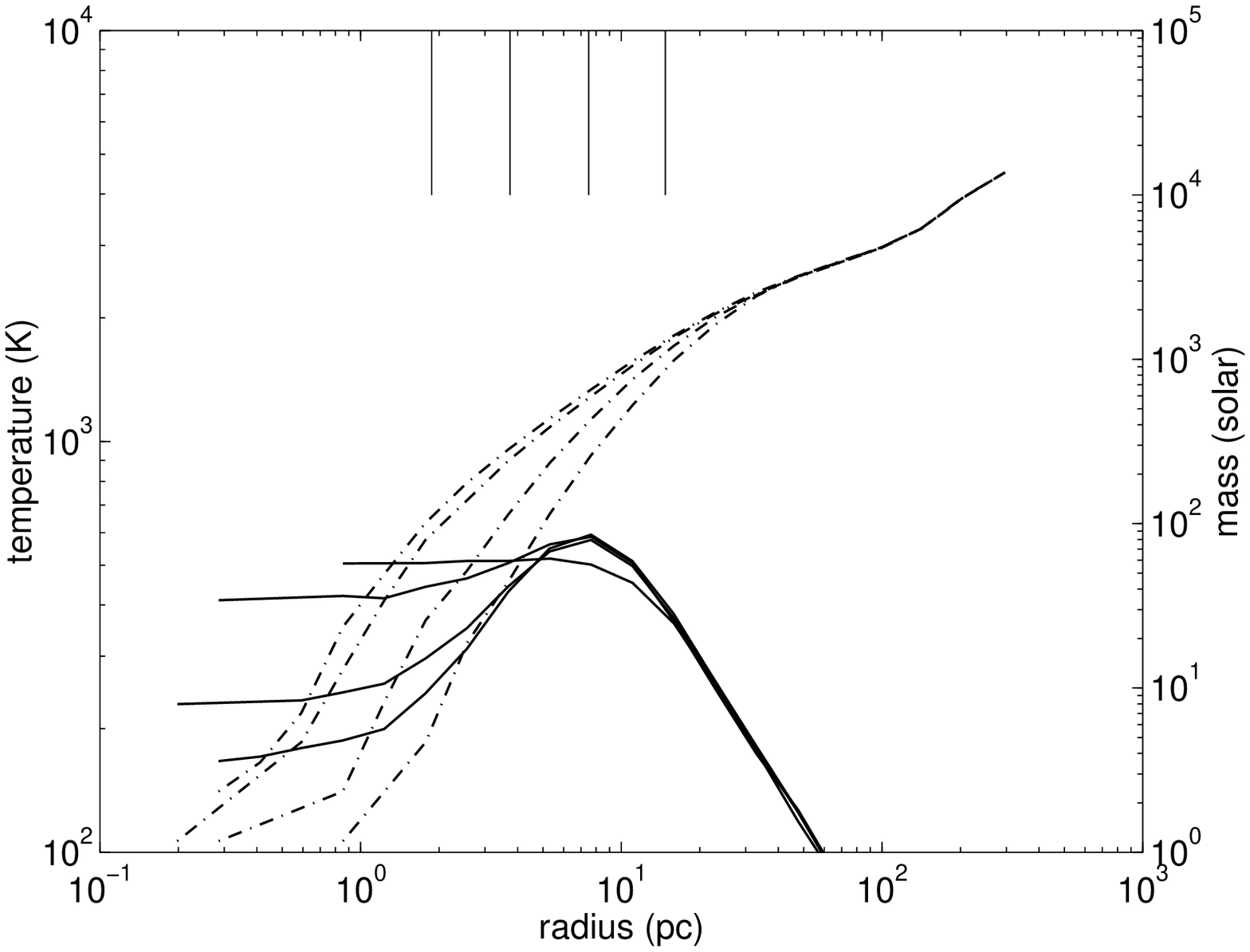}

\caption[Effect of softening on temperature, pressure supported object]
{\label{soft_temperature_supported}Radial profiles of gas temperature (solid
lines) and gas mass (broken lines) for various smoothing lengths (runs 5 to 8)
for an object of total mass of $2\times10^5\,\Msun$ which is not able to cool
efficiently.  As the smoothing length is decreased (vertical lines), the
central temperature decreases and approaches the central temperature
observed in an object that is able to cool efficiently (fig.
\ref{soft_temperature_collapsed}), however the central density is very much
less for pressure supported objects.}

\end{figure}

%% file: collapse.tex
\section{Determination of minimum mass threshold for baryonic condensation}  
\label{ch_collapse}

In CDM-like cosmologies, smaller dark matter perturbations collapse at higher
redshifts, but for the baryonic matter to become self gravitating, enough {\Ht}
must exist for the gas to be able to cool efficiently.  The rate of {\Ht}
formation is related to the baryonic density which is initially driven by the
dark matter potential well.  A minimum mass exists above which the {\Ht}
production rate is sufficient to cool the gas efficiently.  This mass threshold
is a function of redshift, and in this section we explore this parameter space
by simulating top hat perturbations of various masses. Our main interest is to
perform more realistic cosmological simulations of primordial object formation,
and the analytic and top hat simulations provide a reasonable estimate of the
mass scale  at which collapse should occur at a given redshift.  All
simulations presented here are in an Einstein-de Sitter universe with
$\Omega_{dm}=0.94$, $\Omega_{b}=0.06$, and a Hubble constant of
$H_0=65\,\mathrm{km\,s}^{-1}\,\mathrm{Mpc}^{-1}$.  

An object must contain a certain mass of dark matter to be able to force the
baryons into a self gravitating state where star formation may begin.  This 
mass is redshift dependent:  as the universe expands it becomes less dense, and
deeper dark matter gravitational potential wells are required to drive up the
baryonic density to a point where {\Ht} production becomes efficient enough
that the object may cool on a dynamical time.  In this section, using both the
semi-analytic and \nbody methods, we have determined the minimum mass required
for the baryons in an object to become self gravitating $(M_{SG})$ as a
function of redshift.

Figure~\ref{mcollapse} shows the minimum mass threshold versus redshift
relation originally computed by \citet{Tegmark97}.  Using the same method, our
result differs considerably since we have used a different molecular hydrogen
cooling function (see fig.~\ref{Htcoolcurves}), and have included different
chemical reactions.  \citet{Tegmark97}  used the cooling function of
\citet{Hollenbach79}, and assumed that the {\Ht} molecules are all in the para
state and that {\Ht} vibrational cooling was negligible.  At temperatures
higher than $1000\Kelvin$, {\Ht} vibrations may be excited and vibrational
cooling becomes important; this is why the cooling function that
\citet{Tegmark97} used drops below the other curves when $T >1000\Kelvin$. 
More efficient cooling allows smaller, less dense objects to collapse, and
consequently our computation of $M_{SG}(z)$ is a factor of approximately 3 
smaller at $z=50$.  At $z=10$, \citet{Tegmark97} found that the objects able to
cool efficiently had temperatures of approximately $10^4\Kelvin$, and at
this temperature their cooling function is two orders of magnitude lower than
the cooling function we used.  Also, at temperatures greater than
$10^4\Kelvin$ {\Ht} dissociation reactions become important.  Since they
did not incorporate any {\Ht} destruction mechanisms their {\Ht} abundances are
too high, though this is not enough to offset the less efficient cooling
function.  We find that objects able to cool efficiently at $z=10$ are 25 times
less massive.

The results of several $N=2\times 32^3$ top hat simulations are also shown for
comparison with the analytic model (fig.~\ref{mcollapse}).  Inside the top hat
there are $ \simeq 2\times 10^4$ particles, and in cases where gas in the
central region cools quickly, the condensed cores are well resolved and contain
$\simeq 4\times 10^3$ particles.  For a top hat of total initial mass $M_{TH}$
destined to virialize at $z=z_v$,  we have plotted a star if the collapse
succeeded (using the collapse criterion developed in section
\ref{soft_section}) within one dynamical time after virialization and a circle
if the object remained pressure supported.  Some disagreement between the
simulations and the semi-analytic result is to be expected since the latter
assumes that the baryonic overdensity remains constant at $\delta = 200$ after
virialization, and is uniform within the top hat. In spite of the assumptions
made in the semi-analytic result and the different criteria used to determine
if the collapse was successful, the two methods agree to within a factor of 2. 
The slope of the $M_{SG}(z)$ line determined with the semi-analytic method is
slightly steeper than for the simulations but we do not consider this
difference to be of special significance.  Ultimately the precise determination
of $M_{SG}(z)$ is dependent upon the details of the chemical reaction rates and
cooling rates, and the amount of substructure present in the object. Further,
although the threshold for collapse is very sharp because of the
density-cooling feedback (a small increase in mass can cause an appreciable
increase in density), the precise position of the line separating collapse from
pressure support is highly sensitive to small variations in reaction rates and
small numerical changes and hence is simply indicative of the general behaviour
in the $M$--$z$ plane. Our results show that we expect the collapse mass to be
robust within a factor of 2 at a given redshift.

\input{\thisdir/texfigs/mcollapse_vs_z}

%% file: texfigs/mcollapse_vs_z.tex
\begin{figure}
\epsscale{\figscale}
\plotone{\thisdir/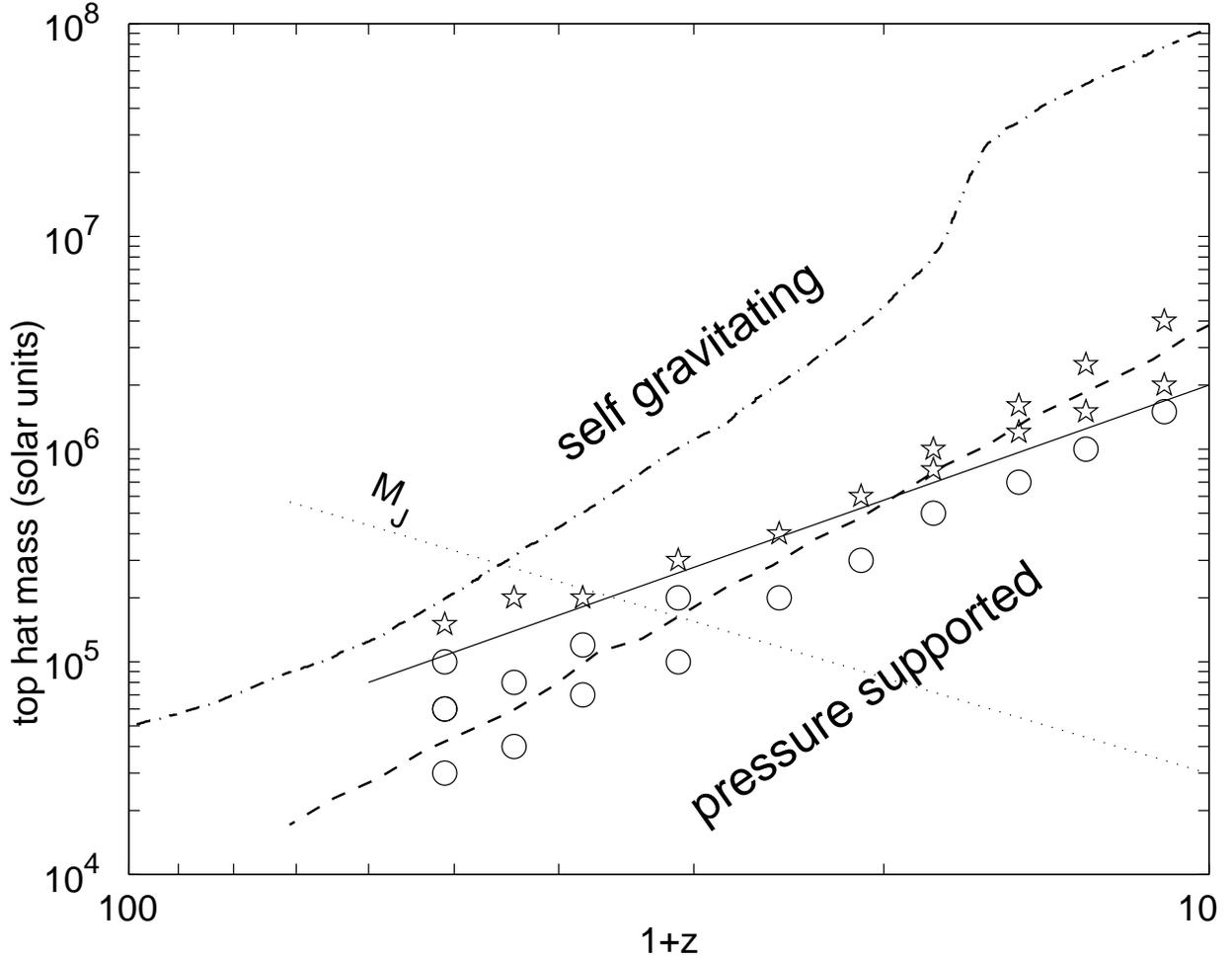}

\caption[Minimum mass threshold versus redshift]{\label{mcollapse}Minimum mass
threshold required for self gravitation ($M_{SG}$) versus redshift.  The
dash-dotted line shows the result of \citet{Tegmark97}, and the dashed line
shows our semi-analytic result.  The dotted line is the Jeans mass.  The
symbols show results of \nbody simulations using $2\times32^3$ particles. 
Objects less massive than $M_{SG}$ at a given redshift will remain pressure
supported (circles); more massive objects cool efficiently and the baryonic
material in the central region becomes self-gravitating (stars).  The solid
line is an approximate fit to the simulation results.}

\end{figure}

%% file: cosmo.tex
\section{Cosmological simulation of first objects}
\label{ch_cosmo}

Our objective here is to follow the growth of objects through merging and 
accretion in the CDM hierarchy and to be able to identify objects that are able
to form enough {\Ht} that they cool efficiently and become self gravitating. Of
primary interest is the determination of the time at which the object exceeds
the Jeans mass threshold, and the time at which the cooling time drops below
the dynamical time.  To follow objects through these phases we simulated a
$(25h^{-1}\,\mathrm{kpc})^3$ volume in an $\Omega=3.4$,
$\Omega_b/\Omega=0.06$, $h=0.65$ universe.  An overdense universe was used
simply to model the  environment of a $3-4\sigma$ perturbation which will lead
to the formation of the first cosmological objects.  The simulation contained
equal numbers of gas and dark matter particles, in total $N=2\times64^3$.  This
gives a mass of 5.1\,\Msun\ for the baryon particles and 81\,\Msun\ for the dark
matter particles.

The mass evolution and merger history of the most massive object in the
simulation is shown in figure~\ref{M200vsz}.  We define $M_{200}$ to be the
mass enclosed by a thin shell that has an overdensity of $200$.   The complete
family tree is very large and to reduce the complexity of the plot, at each
redshift only the direct ancestors of the most massive object is shown.  Lines
connect the smaller objects that have merged into the largest object.    The
separation between the self-gravitating and pressure-supported domains is shown
for both the semi-analytic method and the $N=2 \times 32^3$ top hat
simulations.   The agreement between the cosmological simulation and the top
hat simulation computation of $M_{SG}$ is quite good; the point at which the
object becomes self gravitating is close to the top hat result.

\input{\thisdir/texfigs/M200vsz}

In Figure~\ref{entropy_evol} we plot the entropy evolution of the
approximately 2000 gas particles which reside in the core of the final
cooled object. This figure shows clearly the three stages in
hierarchical evolution. Before the dark matter halo reaches the Jeans
mass, the gas will respond adiabatically to the growing dark matter
potential well, and thus the entropy (defined as $\propto
T/\rho^{2/3}$) will remain constant at the background value. Once the
Jeans mass is reached infalling gas will be shock heated (a
non-adiabatic process) and hence will have its entropy raised. Once
cooling becomes effective, the entropy in the core drops rapidly.

\input{\thisdir/texfigs/entropy_evol}

The baryonic matter in objects less massive than the Jeans mass is pressure
supported against gravitational contraction.  In these objects the baryonic
matter should be relatively unperturbed by the dark matter, and will not have
steep baryonic radial density profiles.  As the baryonic material in an object
approaches the Jeans mass, the baryonic radial density profile will begin to
steepen and become similar in shape to the dark matter radial density
profile.   The most massive object in the simulation exhibits this behaviour,
as shown in figure~\ref{massprofiles}. By $z=28$ the baryon density profile has
started to steepen but its slope is still less than that of the dark matter. 
At $z=24$ the object has exceeded the Jeans mass and the slopes of the baryon
and dark matter density profiles become very close.

\input{\thisdir/texfigs/mass_density_profiles}

\input{\thisdir/texfigs/cooltime}

\input{\thisdir/texfigs/temp_profiles}

At $z=28$, the temperature and {\Ht} fraction (relative to the total hydrogen
abundance) in the most massive object produce a cooling time that is
significantly greater than the dynamical time (fig.~\ref{cooltime}).  Up to
this time, the gas has been relatively mildly perturbed by the growing dark
matter halo. As the dark matter mass in the object increases, the deepening
gravitational potential well drives up the gas density and consequently
accelerates {\Ht} production.  By $z=24$, the {\Ht} abundance and temperature
have increased enough that near the centre of the halo the cooling time drops
below the dynamical time which is $10^7$ years at that time. The cooling  time
is greater than the dynamical time very close to the centre of the object,
however here the gas pressure and temperature are limited by the softening
length, which is approximately 2~pc at this time. The physical state of the
object is such that collapse is inevitable: the dark matter provides a deep
enough potential well, and there is enough molecular hydrogen that the gas
cools efficiently and free falls in the dark matter potential well.  This
increases the gas density and the molecular hydrogen production, thereby
accelerating the {\Ht} cooling, and a runaway feedback mechanism begins.  One
dynamical time later, the core temperature is just above $100\Kelvin$
(fig.~\ref{tempprofiles}), and the baryonic density exceeds the dark matter
density and the object becomes self gravitating (fig.~\ref{massprofiles}).  At
this point the conditions in the cold, dense environment are appropriate for
star formation.

To test for possible numerical resolution effects, we re-sampled the initial
conditions of the $N=2\times64^3$ simulation to produce an $N=2\times32^3$
distribution of particles with similar random density fluctuations.  The radial
profiles of various physical quantities for the most massive object were
compared for both runs and found to agree well.  This convergence indicates that
$2\times 32^3$ particles provides sufficient numerical resolution, at least for
the most massive object. This test suggests that a mass resolution of
40\,\Msun\ per baryonic particle and 650\,\Msun\ per dark matter
particle is sufficient to adequately identify the first self gravitating
clouds in a cosmological simulation. 

Although we have discussed only one object here, we have simulated several
realizations of the same cosmological model to ensure that the object presented
was not anomalous in any way.  Since the two different resolutions discussed
above were found to be convergent, we used the lower resolution and varied only
the random number seed to give different particle positions.  The few
most massive 
objects in these simulations became self gravitating at redshifts ranging from
30 to 15, and all exhibited behaviour similar to that described above: a few
dynamical times after exceeding the Jeans mass, $t_{cool}$ fell below $t_{dyn}$
and the objects became self gravitating within approximately one dynamical
time. Although the redshifts at which the objects became self gravitating
covers a large range, the objects had different masses and all agreed with the
$M_{SG}(z)$ prediction of the top hat simulations to within a factor of two.

Since there exists some uncertainty in the {\Ht} cooling rate and chemical
reaction rates, we investigated the effect on the
collapse of primordial objects of varying key rates.  Recent
determinations of the {\Ht} cooling 
function have differences of a factor $\sim 2$, arising from uncertainty in the
rotational and vibrational {\Hn}-{\Ht} rate coefficients
\citep{Galli98}.  Various authors 
also use different rates for the formation of molecular hydrogen.  We performed
two simulations that differed only in the {\Ht} cooling rate and the rate of
{\Hm} formation (which governs the total {\Ht} abundance).  Inflating
$\Lambda_{\Ht}$ and the {\Hm} formation rate each by a factor of two caused the
most massive object to become self gravitating only slightly earlier ($z=19$
versus $z=17$) and the overall impact on the evolution of the object was quite
small.

By comparing the {\Ht} cooling time scale to the dynamical time, \citet{Tegmark97} argued
that objects will collapse if virial temperatures are high enough to produce a
critical {\Ht} fraction of \mbox{$\sim 5\times 10^{-4}$}.  \citet{Abel97b}
recovered the same abundance in self gravitating regions of objects that they
simulated. However, they argued that the {\Ht} fraction is dependent upon
$\Lambda_{\Ht}$, and that the use of different cooling functions will produce
different values for the {\Ht} abundance threshold needed for collapse. They
stated therefore that their agreement in the critical {\Ht} fraction was
coincidental since they used different molecular hydrogen cooling functions. 
We have also found approximately the same {\Ht} fraction present in the cores
of self gravitating objects (fig.~\ref{cooltime}).  The cooling function that
we have used \citep{Galli98} is quite similar to that used by \citet{Tegmark97} for
$T<400\Kelvin$, so the agreement between the {\Ht} abundances is not
surprising.  The agreement between our work and that of \citet{Abel97b} does
seem unexpected, since the cooling function differs markedly from the one we
employed.  We offer a possible explanation.  A more efficient cooling function
will help to maintain the number of free electrons to catalyze the {\Hm}
channel, but will also retard the production of {\Hm} since the rate falls with
temperature.  These two effects conspire to minimize differences in the {\Ht}
caused by variations in the {\Ht} cooling function.  To investigate this
quantitatively, we performed used several different cooling functions in the
semi-analytic method and recovered each author's results.  Thus, although
different cooling functions will produce different temperatures, the resulting
change in the {\Ht} abundance is small.

%% file: texfigs/M200vsz.tex
\begin{figure}
\epsscale{\figscale}
\plotone{\thisdir/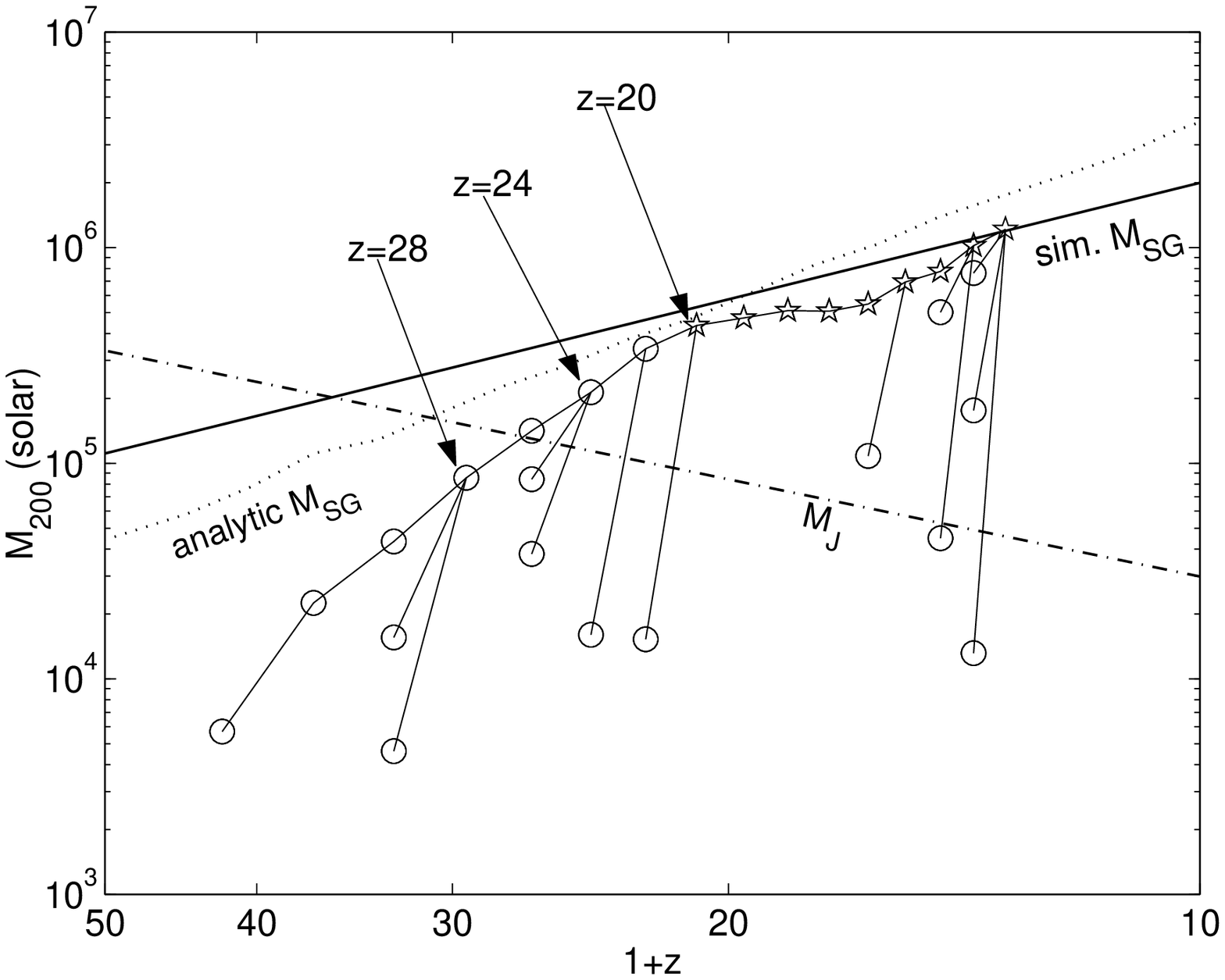}

\caption[Collapse mass threshold]{\label{M200vsz}The mass evolution and merger
history are shown for the most massive object in the cosmological simulation. 
The thin lines connect objects to their predecessors.  The dotted line shows
$M_{SG}$ computed with the semi-analytic method, and the solid line shows the
results from the $N=2\times32^3$ top hat simulations. Initially, the baryonic
matter in the objects is pressure supported (circles), and as mass is accreted
the gas  becomes self gravitating (stars).  Physical properties of the
indicated objects are shown in figures \ref{massprofiles}, \ref{cooltime}, and
\ref{tempprofiles}.  Within $r_{200}$ there are 5100, 2500, and 280 dark matter
particles and 4600, 2100, and 160 gas particles at redshifts of 20, 24, and 28
respectively.}

\end{figure}

%% file: texfigs/entropy_evol.tex
\begin{figure}
\epsscale{\figscale}
\plotone{\thisdir/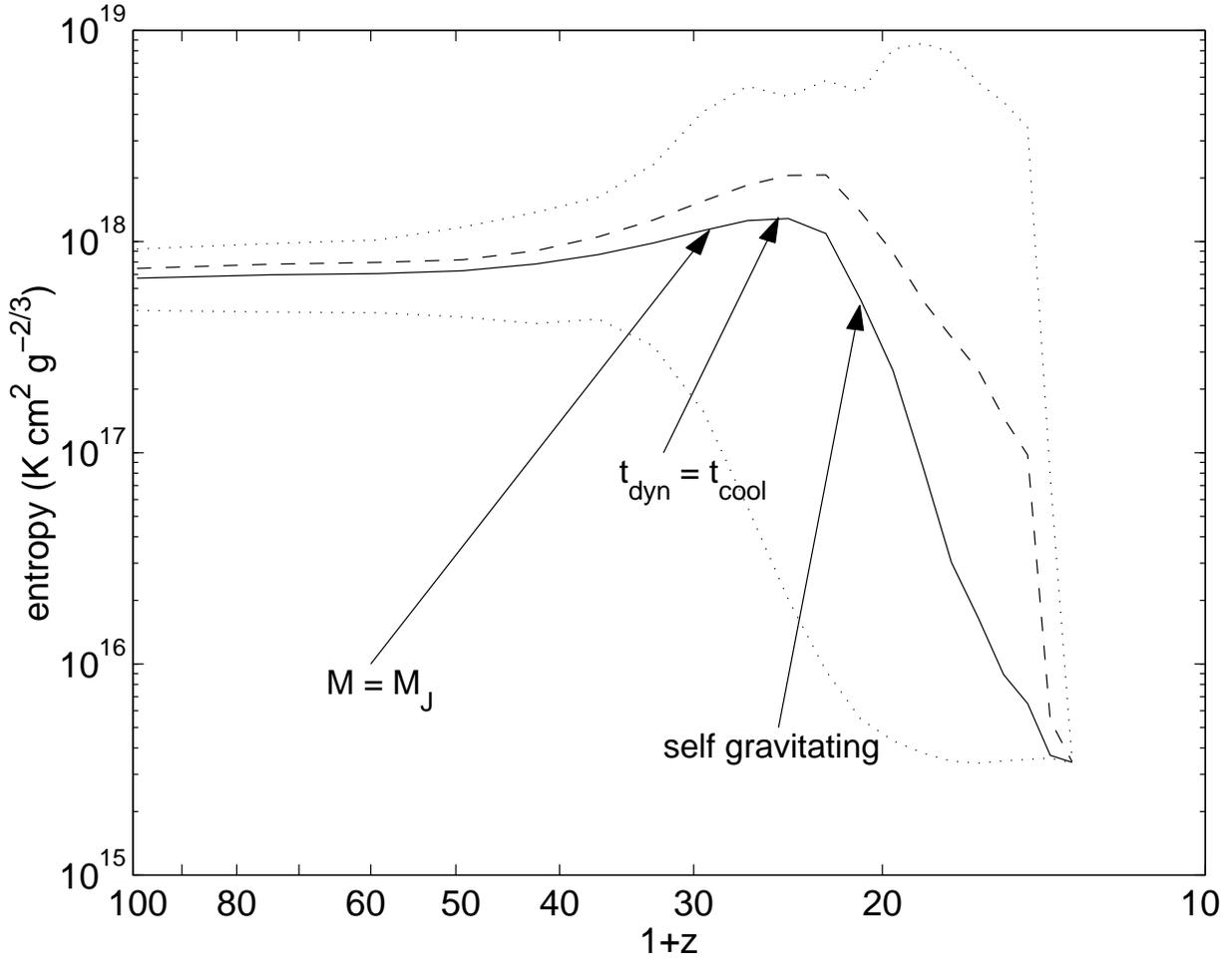}

\caption[Entropy evolution]{\label{entropy_evol}The evolution of the entropy of
the $\sim 2000$ core particles in the most massive object in a cosmological
simulation.  The solid line shows the mean entropy of the core particles,
dotted lines show the particles with the minimum and maximum entropy, and the
dashed line shows one standard deviation above the mean.  Initially the gas is
subject to adiabatic processes only, and the entropy does not change with
time.  As the gas begins to fall into the dark matter potential well and
exceeds the Jeans mass, it is shock heated and the entropy begins to rise. 
After enough {\Ht} has formed, the cooling time falls below the
dynamical time, rapidly reducing the entropy,
and the gas in the core becomes self gravitating within a dynamical time.}

\end{figure}

%% file: texfigs/mass_density_profiles.tex
\begin{figure}
\epsscale{\figscale}
\plotone{\thisdir/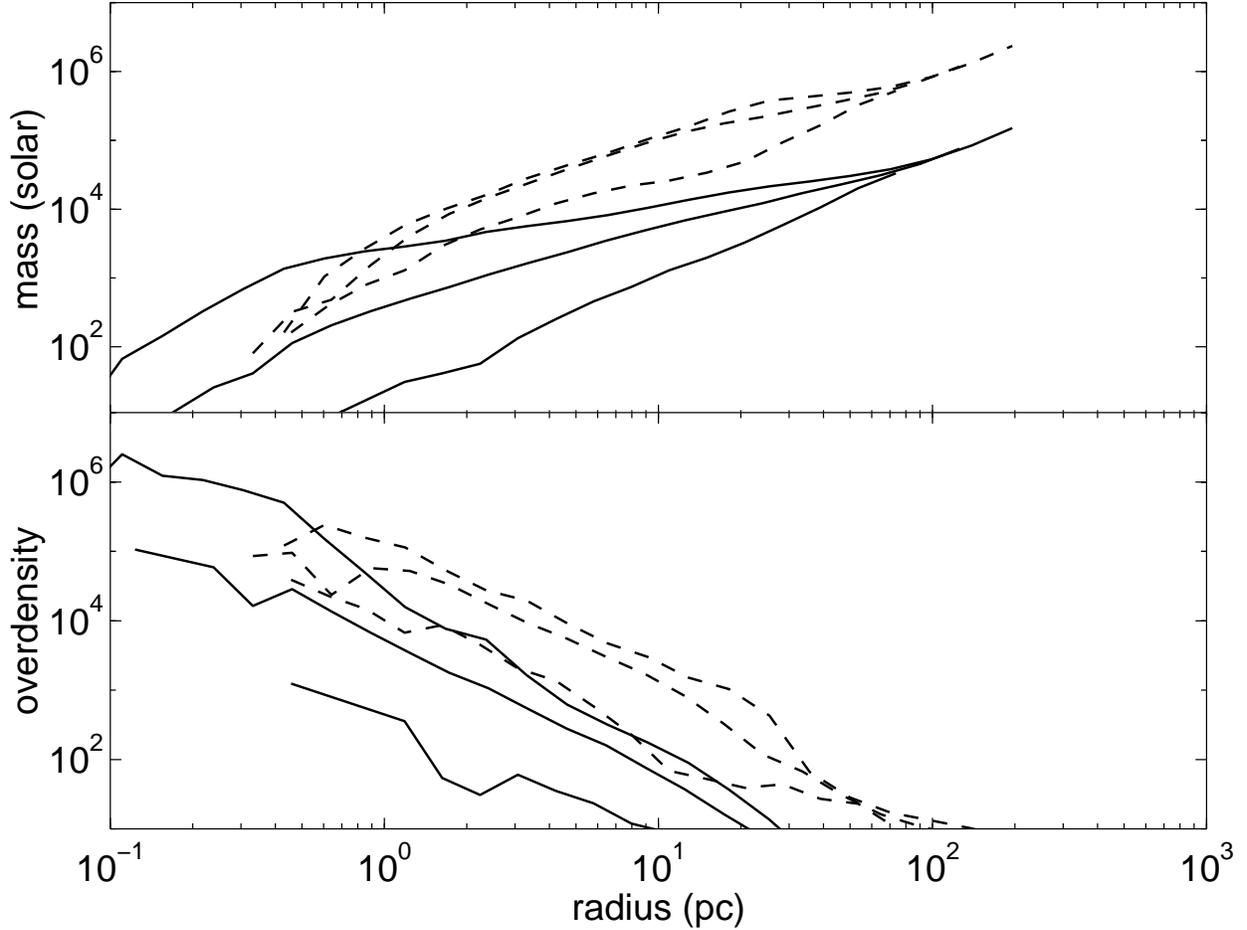}

\caption[Mass and overdensity profiles]{\label{massprofiles}The evolution of
the radial profiles of the mass and overdensity for baryonic (solid lines) and
dark matter (broken lines) are shown for the most massive object in a
cosmological simulation at redshifts of 28, 24, and 20.  These profiles
correspond to the object at the epochs indicated in figure \ref{M200vsz}.  At
$z=24$ enough {\Ht} has formed that the baryons cool efficiently and free fall
in the dark matter potential well.  By $z=20$ --- $10^7$ years or approximately
one dynamical time later --- the object becomes self gravitating.} 

\end{figure}

%% file: texfigs/cooltime.tex
\begin{figure}
\epsscale{\figscale}
\plotone{\thisdir/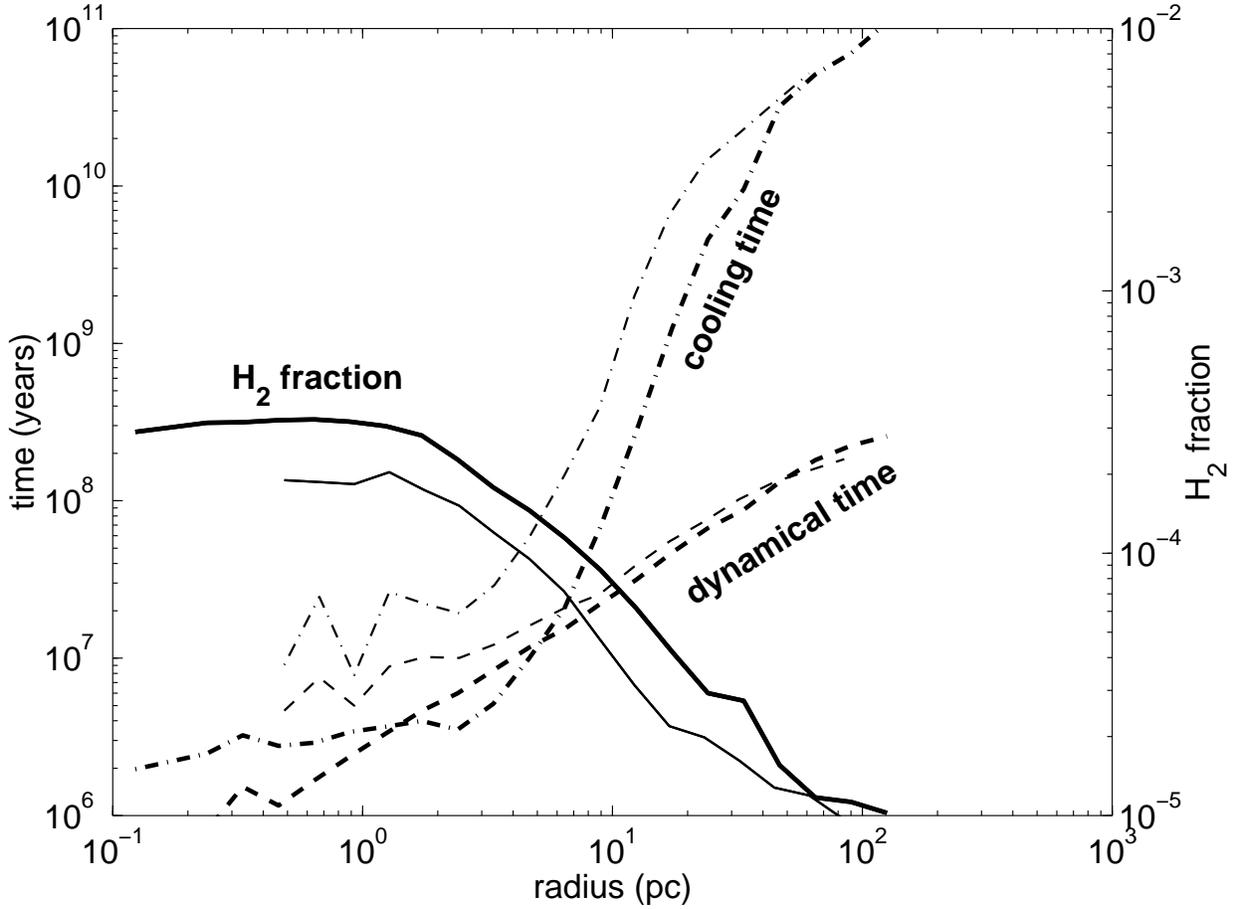}

\caption[Cooling and dynamical times]{\label{cooltime} The cooling time
(dash-dotted lines), dynamical time (dashed lines), and molecular hydrogen
fraction (solid lines) are shown at redshifts of 28 (thin lines) and 24 (thick
lines) for the most massive object in a cosmological simulation.  From $z=28$
to $z=24$ the increase in density and temperature accelerate the {\Ht}
production.  The increase in temperature and {\Ht} abundance conspire to reduce
the cooling time below the dynamical time in part of the central region of the
object by $z=24$.}

\end{figure}

%% file: texfigs/temp_profiles.tex
\begin{figure}
\epsscale{\figscale}
\plotone{\thisdir/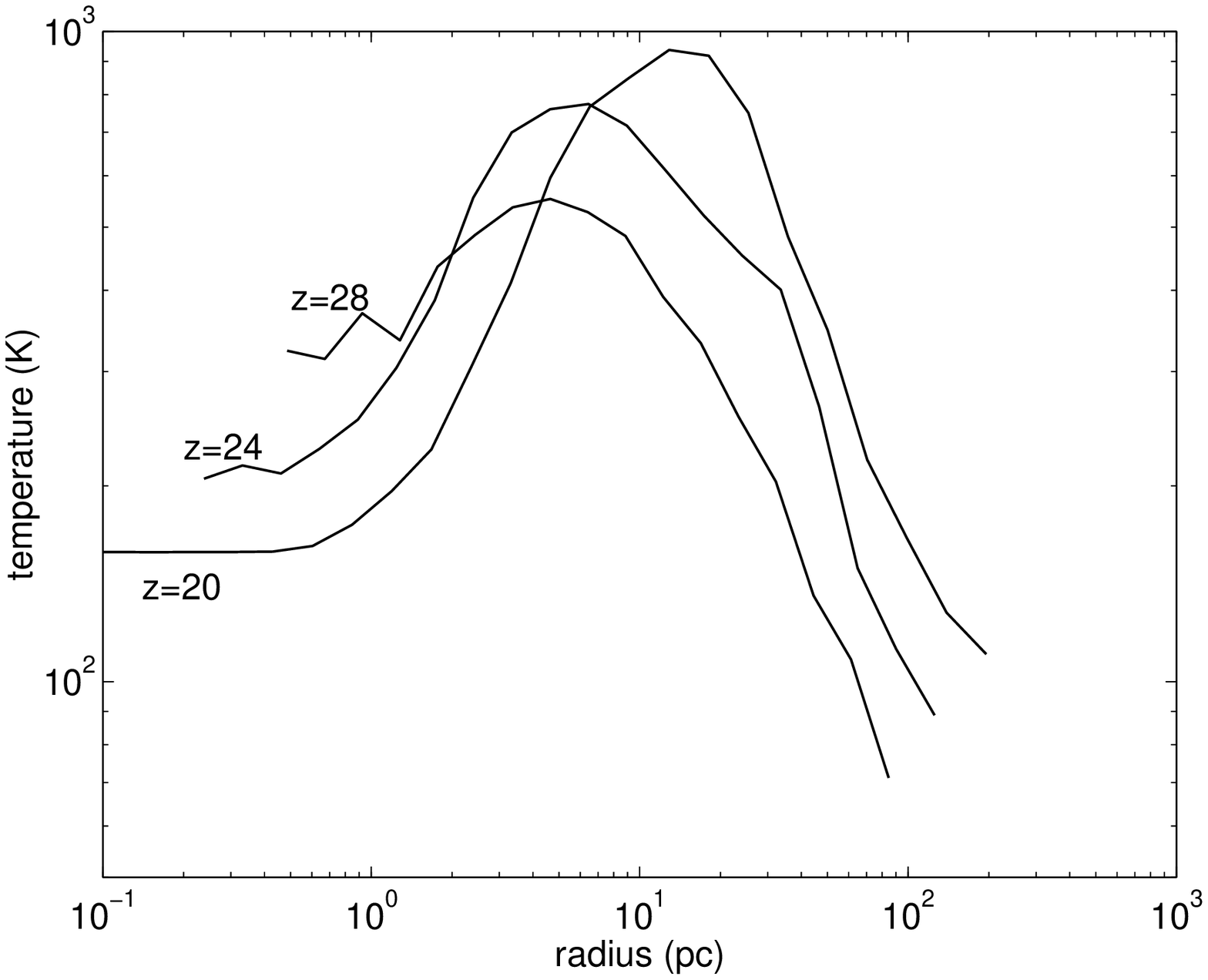}

\caption[Temperature profiles]{\label{tempprofiles}The evolution of the
temperature profile of the baryons in the most massive object in a cosmological
simulation at redshifts of 28, 24, and 20.  These profiles correspond to the
objects indicated in figure \ref{M200vsz}.  The central temperature decreases as
time evolves.  Shock heating of gas falling into the deepening dark
matter potential causes the peak temperature to increase with time, 
and, as the accretion shock moves outward, the peak temperature occurs at
greater distances from the centre of the object.}

\end{figure}

%% file: summary.tex
\section{Summary}
\label{ch_summary}

We have studied the general framework of cooling and condensation of primordial
objects in two component hierarchical cosmologies.  Molecular hydrogen is the
dominant coolant in primordial objects with virial temperatures between
$100\Kelvin$ and $1000\Kelvin$.  Incorporating molecular hydrogen cooling into
a numerical model requires explicit integration of a chemical network because
of the non-equilibrium abundances of several reactants.  The relevant {\Ht}
formation and cooling processes have been added to the cosmological
gravitational and SPH code, Hydra. We have focused on the question of
identifying haloes within which baryons can cool efficiently and in which the
first stars will form, but do not attempt to model the internal structure of
these first clouds in detail.

\citet{Tegmark97} used a simple semi-analytic approach to model the density of
a top hat perturbation, and integrated this along with the chemical equations
and {\Ht} cooling function to determine the evolution of the perturbation. 
This procedure has been followed here with a recently computed {\Ht} cooling
function as a test of the \nbody code. The semi-analytic model approximates
virialization by assuming that the perturbation reaches an overdensity of 200,
remaining at constant physical density thereafter. We find excellent agreement
between the numerical method and the semi-analytic model for cloud collapses in
which the gas remains pressure supported. In cases in which efficient cooling
occurs, the baryon density may substantially exceed the naive analytic value
and the numerical values diverge. If, however, the baryon overdensity at
virialization in the semi-analytic model is adjusted to match the N-body
results, the species abundances come into good agreement once again. We find
that the final, post-virialization, species abundances, densities and
temperatures are insensitive to the high redshift abundance of {\Ht}.

In accordance with the results of \citet{Tegmark97} we find a sharp threshold
(denoted by $M_{SG}$) in the dark matter halo mass--redshift plane separating
haloes in which gas can cool from those in which it cannot. However, our use of
a current cooling function yields $M_{SG}(z)$ values up to an order of
magnitude smaller than found in the earlier work.  The threshold found by the
\nbody method and the semi-analytic method is found to agree to within a factor
of 2 over the redshift range considered.

Our primary purpose is to identify haloes in which the gas may become self
gravitating. For this to occur, the baryons have to condense by a linear factor
$\sim (\Omega/\Omega_b)^{1/3}$, leading to a disparity in the ideal softening
lengths for the dark matter and gas. In order to maintain similar gravitational
and hydrodynamic resolution, and to avoid spurious numerical effects, we do not
allow the softening or hydrodynamic resolution to follow arbitrarily high
baryon densities but instead rely on counting the number of SPH particles
within an SPH smoothing length. We find this technique to be a robust indicator
of whether a halo contains self gravitating gas or not. By this means we are
able to adequately detect the formation of the first self gravitating gas in
CDM-like cosmologies with a gas particle mass of $\sim 40\,\Msun$.

We have performed cosmological simulations of primordial object formation in a
CDM universe.  The evolution and merger histories of several objects were
traced through the simulation and we were able to show that the descriptive
picture of baryon condensation in dark haloes proposed by \citet{White78} is an
accurate representation of the relevant processes leading to the condensation
of gas in dark haloes. We have shown that {\Ht} cooling is very efficient: once
the {\Ht} abundance reaches a level such that the cooling time in the centre of
a halo is less than the local dynamical time, the gas free falls in the
potential well becoming self gravitating one dynamical time later. The point at
which the cooling time first drops below the dynamic time agrees very well with
the minimum mass threshold given by the top hat simulations. The critical {\Ht}
abundance required for collapse, $5\times10^{-4}$, is the same as found by
\citet{Tegmark97}. We find that this abundance is insensitive to the details of
the cooling function used.


%% file: bibliog.tex

%% file: ms.bbl
\begin{thebibliography}{}

\bibitem[\protect\citeauthoryear{Abel et~al.}{Abel et~al.}{1998}]{Abel97b}
Abel, T., Anninos, P., Norman, M.~L.,  \& Zhang, Y. 1998, ApJ,~{508}, 518

\bibitem[\protect\citeauthoryear{Abel et~al.}{Abel et~al.}{1997}]{Abel97a}
Abel, T., Anninos, P., Zhang, Y.,  \& Norman, M.~L. 1997, NewA,~{2}, 181

\bibitem[\protect\citeauthoryear{Anninos \& Norman}{Anninos \&
  Norman}{1996}]{Anninos96}
Anninos, P. \& Norman, M.~L. 1996, ApJ,~{460}, 556

\bibitem[\protect\citeauthoryear{Anninos et~al.}{Anninos
  et~al.}{1997}]{Anninos97}
Anninos, P., Zhang, Y., Abel, T.,  \& Norman, M.~L. 1997, NewA,~{2}, 209

\bibitem[\protect\citeauthoryear{Blumenthal et~al.}{Blumenthal
  et~al.}{1984}]{Blumenthal84}
Blumenthal, G.~R., Faber, S.~M., Primack, J.~R.,  \& Rees, M.~J. 1984,
  Nature,~{311}, 517

\bibitem[\protect\citeauthoryear{Bond et~al.}{Bond et~al.}{1984}]{Bond84}
Bond, J.~R., Arnett, W.~D.,  \& Carr, B.~J. 1984, ApJ,~{280}, 825

\bibitem[\protect\citeauthoryear{Bromm et~al.}{Bromm et~al.}{1999}]{Bromm99}
Bromm, V., Coppi, P.~S.,  \& Larson, R.~B. 1999, ApJ,~{527}, L5

\bibitem[\protect\citeauthoryear{Carr et~al.}{Carr et~al.}{1984}]{Carr84}
Carr, B.~J., Bond, J.~R.,  \& Arnett, W.~D. 1984, ApJ,~{277}, 445

\bibitem[\protect\citeauthoryear{Couchman}{Couchman}{1986}]{Couchman86b}
Couchman, H. M.~P. 1986.
\newblock { Pregalactic activity: some consequences for galaxy formation}.
\newblock Ph.\ D. thesis, King's College, Cambridge

\bibitem[\protect\citeauthoryear{Couchman \& Rees}{Couchman \&
  Rees}{1985}]{Couchman86}
Couchman, H. M.~P. \& Rees, M.~J. 1985, MNRAS,~{221}, 53

\bibitem[\protect\citeauthoryear{Couchman et~al.}{Couchman
  et~al.}{1995}]{Couchman95}
Couchman, H. M.~P., Thomas, P.~A.,  \& Pearce, F.~R. 1995, ApJ,~{452}, 797

\bibitem[\protect\citeauthoryear{Dalgarno \& Lepp}{Dalgarno \&
  Lepp}{1987}]{Dalgarno87}
Dalgarno, A. \& Lepp, S. 1987.
\newblock Chemistry in the early universe.
\newblock In M.~S. Vardya \& S.~P. Tarafdar (Eds.), { Astrochemistry}, Reidel.
  Dordrecht

\bibitem[\protect\citeauthoryear{Dalgarno \& McCray}{Dalgarno \&
  McCray}{1972}]{Dalgarno72}
Dalgarno, A. \& McCray, R.~A. 1972, ARAA,~{10}, 375

\bibitem[\protect\citeauthoryear{Donahue \& Shull}{Donahue \&
  Shull}{1991}]{Donahue91}
Donahue, M. \& Shull, J.~M. 1991, ApJ,~{383}, 511

\bibitem[\protect\citeauthoryear{Dove \& Mandy}{Dove \& Mandy}{1986}]{Dove86}
Dove, J.~E. \& Mandy, M.~E. 1986, ApJL,~{311}, L93

\bibitem[\protect\citeauthoryear{Ferland et~al.}{Ferland
  et~al.}{1992}]{Ferland92}
Ferland, G.~J., Peterson, B.~M., Horne, K., Welsh, W.~F.,  \& Nahar, S.~N.
  1992, ApJ,~{387}, 95

\bibitem[\protect\citeauthoryear{Forrey et~al.}{Forrey et~al.}{1997}]{Forrey97}
Forrey, R.~C., Balakrishnan, N., Dalgarno, A.,  \& Lepp, S. 1997, ApJ,~{489},
  100

\bibitem[\protect\citeauthoryear{Galli \& Palla}{Galli \&
  Palla}{1998}]{Galli98}
Galli, D. \& Palla, F. 1998, AA,~{335}, 403

\bibitem[\protect\citeauthoryear{Gnedin \& Ostriker}{Gnedin \&
  Ostriker}{1997}]{Gnedin97}
Gnedin, N.~Y. \& Ostriker, J.~P. 1997, ApJ,~{486}, 581

\bibitem[\protect\citeauthoryear{Gould}{Gould}{1964}]{Gould64}
Gould, R.~J. 1964, ApJ,~{140}, 638

\bibitem[\protect\citeauthoryear{Gould \& Salpeter}{Gould \&
  Salpeter}{1963}]{Gould63}
Gould, R.~J. \& Salpeter, E.~E. 1963, ApJ,~{138}, 393

\bibitem[\protect\citeauthoryear{Gunn \& Gott}{Gunn \& Gott}{1972}]{Gunn72}
Gunn, J.~E. \& Gott, J.~R. 1972, ApJ,~{176}, 1

\bibitem[\protect\citeauthoryear{Haiman et~al.}{Haiman et~al.}{1999}]{Haiman99}
Haiman, Z., Abel, T.,  \& Rees, M.~J. 1999, astro-ph/9903336

\bibitem[\protect\citeauthoryear{Haiman et~al.}{Haiman et~al.}{1997}]{Haiman97}
Haiman, Z., Rees, M.~J.,  \& Loeb, A. 1997, ApJ,~{476}, 458

\bibitem[\protect\citeauthoryear{Haiman et~al.}{Haiman et~al.}{1996}]{Haiman96}
Haiman, Z., Thoul, A.~A.,  \& Loeb, A. 1996, ApJ,~{464}, 523

\bibitem[\protect\citeauthoryear{Hill \& Silk}{Hill \& Silk}{1975}]{Hill75}
Hill, J.~K. \& Silk, J. 1975, ApJL,~{202}, L97

\bibitem[\protect\citeauthoryear{Hindmarsh}{Hindmarsh}{1983}]{Hindmarsh83}
Hindmarsh, A.~C. 1983.
\newblock In R.~S. Stepleman \& M.~Carver (Eds.), { Scientific Computing},
  Amsterdam. North-Holland

\bibitem[\protect\citeauthoryear{Hirasawa}{Hirasawa}{1969}]{Hirasawa69}
Hirasawa, T. 1969, Progress of Theoretical Physics,~{42}, 523

\bibitem[\protect\citeauthoryear{Hollenbach \& McKee}{Hollenbach \&
  McKee}{1979}]{Hollenbach79}
Hollenbach, D. \& McKee, C.~F. 1979, ApJS,~{41}, 555

\bibitem[\protect\citeauthoryear{Hutchins}{Hutchins}{1976}]{Hutchins76}
Hutchins, J.~B. 1976, ApJ,~{205}, 103

\bibitem[\protect\citeauthoryear{Janev et~al.}{Janev et~al.}{1987}]{Janev87}
Janev, R.~K., Langer, W.~D.,  \& Evans, W.~D. 1987.
\newblock { Elementary Processes in Hydrogen-Helium Plasmas}.
\newblock Berlin: Springer-Verlag

\bibitem[\protect\citeauthoryear{Karpas et~al.}{Karpas et~al.}{1979}]{Karpas79}
Karpas, Z., Anicich, V.,  \& Huntress 1979, J. Chem. Phys.,~{70}, 2877

\bibitem[\protect\citeauthoryear{Kashlinsky \& Rees}{Kashlinsky \&
  Rees}{1983}]{Kashlinsky83}
Kashlinsky, A. \& Rees, M.~J. 1983, MNRAS,~{205}, 955

\bibitem[\protect\citeauthoryear{Knox}{Knox}{1998}]{Knox98}
Knox, L. 1998, Phys. Rev. Lett.,~{81}, 2004

\bibitem[\protect\citeauthoryear{Kwan}{Kwan}{1977}]{Kwan77}
Kwan, J. 1977, ApJ,~{216}, 713

\bibitem[\protect\citeauthoryear{Lepp \& Shull}{Lepp \& Shull}{1983}]{Lepp83}
Lepp, S. \& Shull, J.~M. 1983, ApJ,~{270}, 578

\bibitem[\protect\citeauthoryear{Lepp \& Shull}{Lepp \& Shull}{1984}]{Lepp84}
Lepp, S. \& Shull, J.~M. 1984, ApJ,~{280}, 465

\bibitem[\protect\citeauthoryear{Loeb}{Loeb}{1997}]{Loeb97}
Loeb, A. 1997.
\newblock The first stars and quasars in the universe.
\newblock In E.~P. Smith \& A.~Koratkar (Eds.), { ASP Conference Series -
  Science with the NGST}, Volume 133, pp.\ ~73

\bibitem[\protect\citeauthoryear{Martin et~al.}{Martin et~al.}{1996}]{Martin96}
Martin, P.~G., Schwarz, D.~H.,  \& Mandy, M.~E. 1996, ApJ,~{461}, 265

\bibitem[\protect\citeauthoryear{Meiksin et~al.}{Meiksin
  et~al.}{1999}]{Meiksin99}
Meiksin, A., White, M.,  \& Peacock, J.~A. 1999, MNRAS,~{304}, 851

\bibitem[\protect\citeauthoryear{Omukai \& Nishi}{Omukai \&
  Nishi}{1998}]{Omukai98}
Omukai, K. \& Nishi, R. 1998, ApJ,~{508}, 141

\bibitem[\protect\citeauthoryear{Osterbrock}{Osterbrock}{1974}]{Osterbrock74}
Osterbrock, D.~E. 1974.
\newblock { Astrophysics of Gaseous Nebulae}.
\newblock San Francisco: Freeman and Co.

\bibitem[\protect\citeauthoryear{Peebles}{Peebles}{1993}]{Peebles93}
Peebles, P. J.~E. 1993.
\newblock { Principles of Physical Cosmology}.
\newblock Princeton: Princeton University Press

\bibitem[\protect\citeauthoryear{Peebles \& Dicke}{Peebles \&
  Dicke}{1968}]{Peebles68}
Peebles, P. J.~E. \& Dicke, R.~H. 1968, ApJ,~{154}, 891

\bibitem[\protect\citeauthoryear{Press et~al.}{Press et~al.}{1992}]{Press92}
Press, W.~H., Teukolsky, S.~A., Vetterling, W.~T.,  \& Flannery, B.~P. 1992.
\newblock {\em Numerical Recipes in Fortran\/} (2 ed.).
\newblock Cambridge: Cambridge University Press

\bibitem[\protect\citeauthoryear{Puy et~al.}{Puy et~al.}{1993}]{Puy93}
Puy, D., Alecian, G., {Le Bourlot}, J., L\'{e}orat, J.,  \& {Pineau des
  For\^{e}ts}, G. 1993, AA,~{267}, 337

\bibitem[\protect\citeauthoryear{Rapp \& Francis}{Rapp \&
  Francis}{1962}]{Rapp62}
Rapp, D. \& Francis, W.~E. 1962, J. Chem. Phys,~{37}, 2631

\bibitem[\protect\citeauthoryear{Schneider et~al.}{Schneider
  et~al.}{1994}]{Schneider94}
Schneider, I.~F., Dulieu, O., Giusti-Suzor, A.,  \& Roueff, E. 1994,
  ApJ,~{424}, 983

\bibitem[\protect\citeauthoryear{Shapiro \& Kang}{Shapiro \&
  Kang}{1987}]{Shapiro87}
Shapiro, P.~R. \& Kang, H. 1987, ApJ,~{318}, 32

\bibitem[\protect\citeauthoryear{Shaver et~al.}{Shaver et~al.}{1999}]{Shaver99}
Shaver, P.~A., Windhorst, R.~A., Madau, P.,  \& de~Bruyn, A.~G. 1999,
  AA,~{345}, 380

\bibitem[\protect\citeauthoryear{Solomon \& Werner}{Solomon \&
  Werner}{1971}]{Solomon71}
Solomon, P.~M. \& Werner, M.~W. 1971, ApJ,~{165}, 41

\bibitem[\protect\citeauthoryear{Stahler}{Stahler}{1986}]{Stahler86}
Stahler, S.~W. 1986, PASP,~{98}, 1081

\bibitem[\protect\citeauthoryear{Stancil}{Stancil}{1994}]{Stancil94}
Stancil, P.~C. 1994, ApJ,~{430}, 360

\bibitem[\protect\citeauthoryear{Tegmark et~al.}{Tegmark
  et~al.}{1997}]{Tegmark97}
Tegmark, M., Silk, J., Rees, M.~J., Abel, T.,  \& Palla, F. 1997, ApJ,~{474}, 1

\bibitem[\protect\citeauthoryear{Thacker et~al.}{Thacker
  et~al.}{1998}]{Thacker99}
Thacker, R.~J., Tittley, E.~R., Pearce, R.~R.,  \& Couchman, H. M.~P. 1998,
  astro-ph/9809221

\bibitem[\protect\citeauthoryear{White \& Rees}{White \& Rees}{1978}]{White78}
White, S. D.~M. \& Rees, M.~J. 1978, MNRAS,~{183}, 341

\bibitem[\protect\citeauthoryear{Wishart}{Wishart}{1979}]{Wishart79}
Wishart 1979, MNRAS,~{187}, 59P

\end{thebibliography}
